\documentclass[journal=jacsat,manuscript=article,layout=twocolumn]{achemso}
\usepackage[fontsize=10pt]{fontsize}

\usepackage[version=4]{mhchem} 
\usepackage{times}
\usepackage{upgreek}
\usepackage{siunitx}
\usepackage{graphicx}
\usepackage{lscape}
\usepackage{booktabs}
\usepackage{multirow}
\usepackage[flushleft]{threeparttable}
\usepackage{amsmath,amssymb}
\usepackage{natbib}
\usepackage{comment}
\usepackage{hyperref}

\SectionNumbersOn

\author{Gabi Wenzel}
\email{gwenzel@mit.edu}
\affiliation[MIT]
{Department of Chemistry, Massachusetts Institute of Technology, Cambridge, MA 02139, USA}
\author{Martin S. Holdren}
\email{holdrenm@mit.edu}
\affiliation[MIT]
{Department of Chemistry, Massachusetts Institute of Technology, Cambridge, MA 02139, USA}
\author{D. Archie Stewart}
\affiliation[MIT]
{Department of Chemistry, Massachusetts Institute of Technology, Cambridge, MA 02139, USA}
\author{Hannah Toru Shay}
\affiliation[MIT]
{Department of Chemistry, Massachusetts Institute of Technology, Cambridge, MA 02139, USA}
\author{Alex N. Byrne}
\affiliation[MIT]
{Department of Chemistry, Massachusetts Institute of Technology, Cambridge, MA 02139, USA}
\author{Ci Xue}
\affiliation[MIT]
{Department of Chemistry, Massachusetts Institute of Technology, Cambridge, MA 02139, USA}
\author{Brett A. McGuire}
\email{brettmc@mit.edu}
\affiliation[MIT]
{Department of Chemistry, Massachusetts Institute of Technology, Cambridge, MA 02139, USA}
\alsoaffiliation[NRAO]
{National Radio Astronomy Observatory, Charlottesville, VA 22903, USA}

  \title{Laboratory rotational spectra of cyanocyclohexane and its siblings (1- and 4-cyanocyclohexene) using a compact CP-FTMW spectrometer for interstellar detection}


\begin{document}

\begin{tocentry}




\includegraphics[width=8cm]{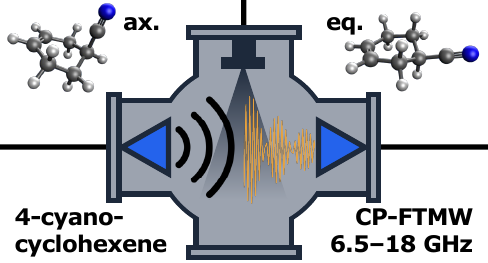}

\end{tocentry}

\begin{abstract}

Chirped-pulse Fourier transform microwave (CP-FTMW) spectroscopy is a versatile technique to record broadband gas-phase rotational spectra, enabling detailed investigations of molecular structure, dynamics, and hyperfine interactions. Here, we present the development and application of a CP-FTMW spectrometer operating in the $6.5-18\,\mathrm{GHz}$ frequency range, studying cyanocyclohexane, 1-cyanocyclohexene, and 4-cyanocyclohexene using a heated pulsed supersonic expansion source. The dynamic range, experimental resolution, and high sensitivity enable observation of multiple conformers, precise measurements of hyperfine splitting arising from nuclear quadrupole coupling due to the nitrogen atom in the cyano group, as well as the observation of {singly \ce{^13C}- and \ce{^15N}-substituted isotopic isomers} in natural abundance. Using the latter, precise structures for the molecules are derived. The accurate rotational spectra enabled a search for these species toward the dark, cold molecular cloud TMC-1; no signals are found, and we discuss the implications of derived upper limits on the interstellar chemistry of the {cyanocyclohexane} family.

\end{abstract}


\section{Introduction}

While more than 300 molecules have now been detected in the interstellar medium (ISM)~\cite{mcguire_2021_2022}, the detection of benzonitrile (\ce{c-C6H5CN}), a nitrogen-bearing aromatic hydrocarbon, in the Taurus molecular cloud TMC-1~\cite{mcguire_detection_2018}, sparked special interest in the astrochemistry community due to the hypothesized role of conjugated, cyclic species in astrobiology and in polycyclic aromatic hydrocarbon (PAH) formation. Indeed, since the detection of benzonitrile, nine PAHs have been identified in TMC-1 using radio observations from facilities such as the 100-m Robert C. Byrd Green Bank Telescope (GBT) and the Yebes 40-m radio telescope~\cite{burkhardt_discovery_2021,cernicharo_pure_2021,sita_discovery_2022,mcguire_detection_2021,cernicharo_discovery_2024,wenzel_detection_2024,wenzel_detections_2025}.

These detections were only possible due to the interplay between laboratory rotational spectroscopy and radio astronomy. Both exploit the same quantum mechanical behavior of molecular rotation, releasing energy corresponding to transitions between rotational states of an excited molecule and thus giving rise to characteristic line features or ``fingerprints''. As such, rotational spectroscopy has proven a critical tool for astrochemists, with active, ongoing research in millimeter, microwave, and terahertz regimes.  

In the late 1970s, Balle and Flygare developed the molecular beam {Fabry–Perot} Fourier-transform microwave (FTMW) spectrometer~\cite{balle_new_1979,balle_new_1980,balle_fabryperot_1981}. The combination of the pulsed molecular beam---which allows for rapid cooling and thus the ability to study large molecules and clusters---with the powerful amplification provided by the {Fabry-Perot} microwave cavity changed the landscape of the field. Cavity-enhanced FTMW could simultaneously achieve high frequency resolution while increasing the candidate pool of species to radicals, ions, and complexes due to its enhanced sensitivity. 

{However, due to the narrowband nature of the Fabry-Perot cavity used in the Balle-Flygare FTMW, in order to obtain a broadband spectrum, the cavity must be tuned to specific resonances only $\sim$0.5\,MHz wide or lower before moving onto the next.} While many modern systems perform this piecewise acquisition automatically, the total time required to acquire a broadband scan can remain quite long. First introduced in 2008 by \citet{brown_broadband_2008}, the broadband chirped-pulse (CP) FTMW technique addressed this issue by taking advantage of technological advancements in spectrometer components like arbitrary waveform generators and fast oscilloscopes. These instruments typically offer more than an order of magnitude increase in acquisition speed at the cost of a modest decline in sensitivity and spectral resolution, and have proven successful in extensive mixture analysis leading to interstellar detections of complex organic molecules~\cite{park_perspective_2016,mccarthy_exhaustive_2020,loru_detection_2023,fried_rotational_2024}.

Quantum chemical calculations have also improved, allowing one to predict the rotational spectrum of a molecule to high accuracy through computationally inexpensive geometry optimizations~\cite{melli_extending_2021, yazidi_ab_2019, lee_bayesian_2020}. A level of theory commonly used for these calculations is the density functional theory (DFT) B3LYP method with added D3(BJ) empirical dispersion correction and using the 6-311G++(d,p) basis set
\cite{becke_densityfunctional_1993, grimme_consistent_2010, grimme_effect_2011, grimme_effects_2013}. For molecules of the size and nature of those commonly searched for in the ISM, this level of theory works well at providing accurate structures while being computationally inexpensive~\cite{marshall_rotational_2017, neill_analysis_2023}. This typically yields rotational constants with (often substantially) better than $2\,\%$ errors compared to their experimental fits~\cite{neill_analysis_2023}. However, this level of accuracy has been shown to not be sufficient for interstellar detections purely from calculations, where a fractional error on the rotational constants of ${\lesssim}1.5\,\mathrm{ppm}$ is required~\cite{wenzel_detection_2024,barone_hunting_2023} for all but a handful of (typically linear) species~\cite{cernicharo_magnesium_2023,cernicharo_quijote_2024}. Therefore, the measurement of the species of interest in the laboratory is required to refine calculated rotational constants to sufficiently high accuracy for an interstellar search.

\begin{figure}[t!]
    \centering
    \includegraphics[width=\columnwidth]{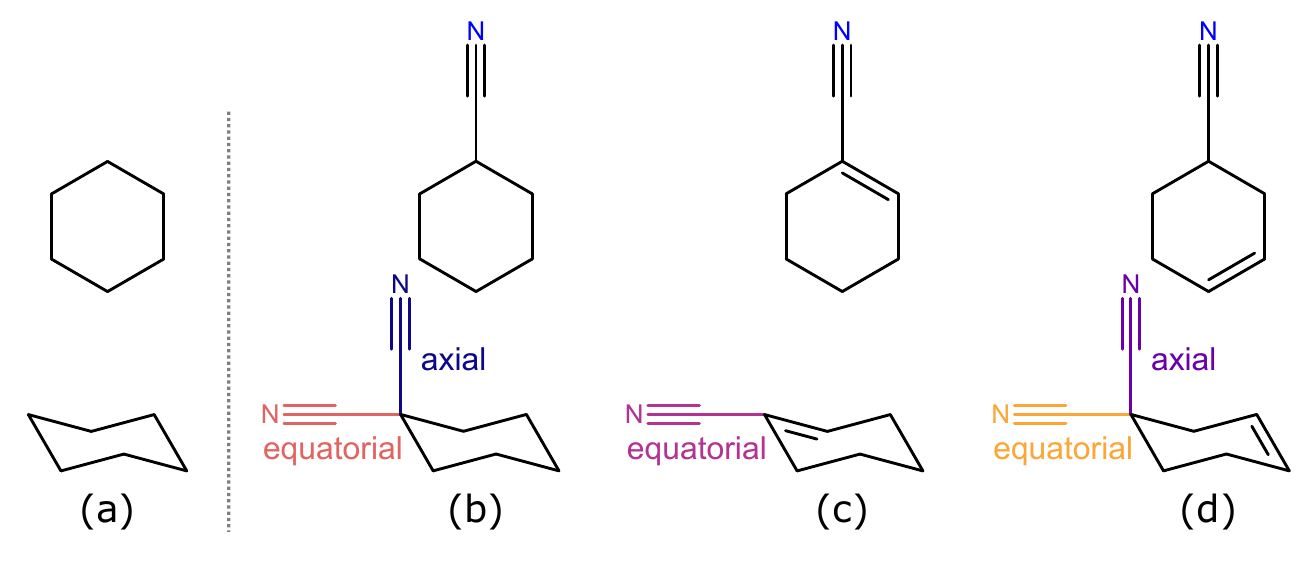}
    \caption{2D molecular structures of the studied species. (a) Unsubstituted cyclohexane depicted from top and in side view in its chair conformation. Analogously, (b) through (d) correspond to cyanocyclohexane, 1-cyanocyclohexene, and 4-cyanocyclohexene, respectively. Their possible chair-axial and chair-equatorial conformations are shown and color-coded throughout the manuscript.}
    \label{fig:mol_structures2d}
\end{figure}

Here, we describe the design and performance of a new compact CP-FTMW spectrometer operating a frequency range of $6.5-18\,\mathrm{GHz}$ using part of the hardware of the portable cavity instrument from F. Lovas' instrument at NIST~\cite{suenram_portable_1999}. This frequency regime is extremely well-suited for astrochemical studies of small aromatic species, whose strongest transitions at low temperatures peak at these frequencies. To demonstrate the capabilities of this new instrument, we investigated the broadband rotational spectra of cyanocyclohexane, 1-cyanocyclohexene, and 4-cyanocyclohexene (see Fig.~\ref{fig:mol_structures2d} for their structures), which share structural similarities with benzonitrile and are appealing targets for potential astronomical detection. We describe the instrument in detail and benchmark it with the previously studied carbonyl sulfide, cyanocyclohexane, and 1-cyanocyclohexene. Further, we present the first rotational spectroscopic analysis of 4-cyanocyclohexene in its two conformations, including all {singly-substituted heavy atom isotopic isomers observed in natural abundance,} and derive their precise structures. Lastly, we conduct an astronomical search for these species toward TMC-1, and discuss the chemical implications of their non-detections.

\section{Methods}

\subsection{Compact CP-FTMW Spectrometer}

\begin{figure}[t!]
    \centering
    \includegraphics[width=\columnwidth]{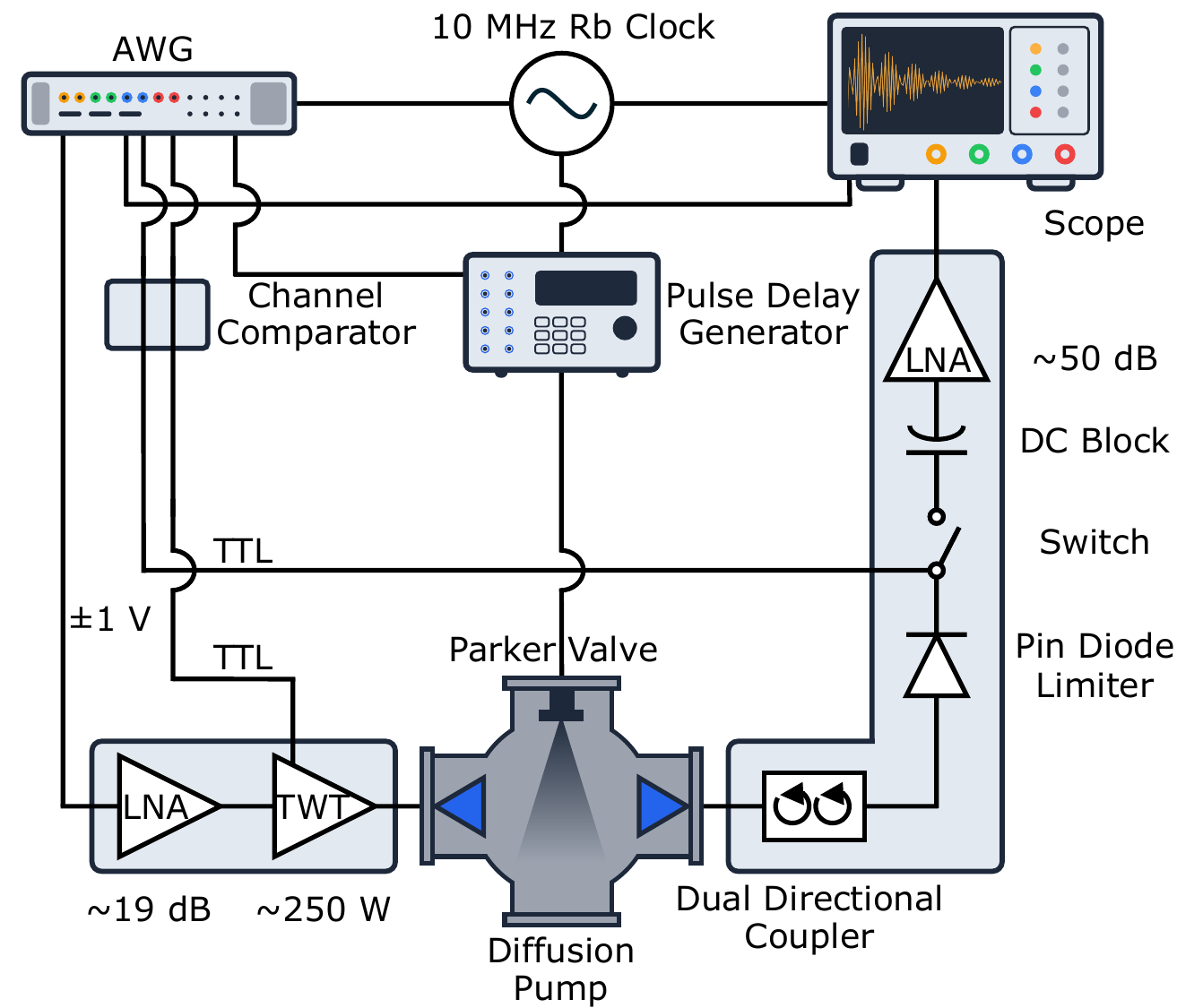}
    \caption{Schematic block diagram of the CP-FTMW spectrometer operating from $6.5$ to $18.0\,\mathrm{GHz}$. A schematic block diagram and list detailing each part involved in the circuit can be found in the SI in Fig.~\ref{fig:SI_tweety_circuit} and Table~\ref{tab:parts}, respectively.}
    \label{fig:tweety_circuit}
\end{figure}

In the following, our new CP-FTMW spectrometer is described and a schematic block diagram is depicted in Fig.~\ref{fig:tweety_circuit}. All parts are listed and diagrammed in the Supporting Information (SI) in Table~\ref{tab:parts} and Fig.~\ref{fig:SI_tweety_circuit}, respectively. A linear frequency-swept chirped pulse covering the $6.5-18.0\,\mathrm{GHz}$ frequency range is generated by an arbitrary waveform generator (AWG, Keysight M8195A, 4 channels) at a sampling rate of $65\,\mathrm{GS/s}$. The chirp is output on channel 1 and sent through a low noise amplifier (LNA, RF Lambda R06G18GSA, ${\sim}19\,\mathrm{dB}$) before reaching the traveling wave tube (TWT) amplifier (Applied Systems Engineering, 367X/Ku, ${\sim}250\,\mathrm{W}$). The high-power TWT output is broadcast into an Eccosorb-covered six-cross compact ISO160--200 vacuum chamber using a standard gain horn antenna (Microtech HWRD650). The chamber is evacuated using a diffusion pump (Edwards Diffstak 160) and a rotary vane pump (Edwards RV18), ultimately reaching pressures in the low $10^{-6}\,\mathrm{mbar}$.

The target molecules, seeded in inert gas, are introduced using a modified pulsed solenoid valve (Parker Series 9) equipped with a heated sample reservoir (see Section~\ref{sec:nozzle} and Fig.~\ref{fig:Heated-sample-holder} in the SI and similar builds in the literature~\cite{sedo_rotational_2019, suenram_reinvestigation_2001, schmitz_multi-resonance_2012, hazrah_structure_2022, cabezas_laboratory_2023, thorwirth_rotational_2005, neill_online_2019}), located approximately $13\,\mathrm{cm}$ from the center of the chamber, creating a jet-cooled molecular beam with a rotational temperature ($T_\mathrm{rot}$) of approximately $2\,\mathrm{K}$ on the orthogonal axis of the chirp excitation. After excitation of the molecules in the supersonic expansion, the molecular free-induction decay (FID) is collected by a second horn antenna, whose face is located $13\,\mathrm{cm}$ from the center of the chamber. The circuit on the receiver side consists of a dual directional coupler (RF Lambda RFDDC2G18G30), a pin diode limiter (Pasternack PE80L3000, $200\,\mathrm{W}$), a single pole single throw (SPST) switch (RF Lambda RFSPSTA0118G), and a DC block (RF Lambda RFDCBLK26SMA), all to protect the ultra low noise amplifier (RF Lambda RLNA06G18G45) from the high-power TWT amplifier output. The amplified molecular FID is then digitized on channel 1 of a fast oscilloscope (Keysight DSOV204A) with a sampling rate of $80\,\mathrm{GS/s}$ to cover $20\,\mathrm{GHz}$ bandwidth.

Timing of the experimental cycle is controlled by a delay pulse generator (Quantum Composer 9528) which is coupled to a $10\,\mathrm{MHz}$ signal generated by a Rb clock (Stanford Research Systems FS725) and modulated with a step function created by a synthesized function generator (Stanford Research Systems DS345, not depicted in Fig.~\ref{fig:tweety_circuit}). The AWG and oscilloscope are phase-locked and referenced to the same $10\,\mathrm{MHz}$ Rb clock. The solenoid valve is powered using a pulse driver (Parker Iota One, not shown in Fig.~\ref{fig:tweety_circuit}) which in turn is triggered by the delay pulse generator. As well, upon receiving a trigger signal from the delay pulse generator, the AWG will output the chirp on channel 1, along with two marker pulses on channel 3 and 4. These trigger the operation of the SPST switch, oscilloscope (both channel 3), and TWT (channel 4), and account for opening and closing times in the switch and internal ramp-up and ramp-down times in the TWT, respectively. To be operable as triggers for the switch and TWT, these AWG marker outputs are amplified by a channel comparator (Pulse Research Lab PRL-350TTL) to a standard TTL pulse of approximately $5\,\mathrm{V}$.

Our standard mode of operation is probing each supersonic expansion gas pulse (300--600\,$\mathrm{\upmu s}$ in duration) with 10 to 20 chirps fired at 12--20\,$\mathrm{\upmu s}$ intervals. The length of the chirp can be modified to optimize excitation conditions for each molecular species studied (typically 300--3000\,$\mathrm{ns}$ in duration). The resulting FIDs are collected by the fast oscilloscope in segmented mode and transferred via a fast switch ($1\,\mathrm{GB/s}$) to the lab computer. Using the open-source data acquisition program for CP-FTMW spectrometers, \textsc{blackchirp}\footnote{\url{https://github.com/kncrabtree/blackchirp}}, we average the FIDs, probing the same region of each gas pulse, into one ``frame," and finally average all frames into one final FID. Operating the spectrometer in this mode and cycling at 3--5\,Hz yields effective experimental repetition rates of 30--60\,Hz, only limited by our decreased pumping speed of $700\,\mathrm{l/s}$ in the compact setup (when comparing to larger setups using diffusion pumps with speeds up to $4500\,\mathrm{l/s}$) and data transfer rates which strongly depend on the length of FID (10--20\,$\mathrm{\upmu s}$). 

All FIDs presented in this work were processed identically: truncated to their first $10\,\mathrm{\upmu s}$, zero-padded to a length of $2^{22}$ bins, and filtered with a Hanning window before their Fast Fourier transform (FFT) was computed. The FIDs are padded to a power of 2 to maximize the efficiency of the FFT algorithm, and a length of $2^{22}$ bins was specifically chosen for its balance between spectral resolution and the size of the spectrum files. Using \textsc{blackchirp}, the FID processing is performed live and the resulting spectra can be inspected online during the experiment. The fast oscilloscope records the FID as voltage vs. time. We then perform a FFT and normalize such that the root-mean-square (RMS) noise is set to 1, yielding spectra in signal-to-noise ratio (SNR) vs. frequency.

\subsection{Quantum Chemical Calculations}

The 2D structures of the molecules studied in this work are shown in Fig.~\ref{fig:mol_structures2d}, and their {equilibrium} geometries were first optimized using density functional theory (DFT) as part of the Gaussian16 suite of programs~\cite{m_j_frisch_g_w_trucks_h_b_schlegel_g_e_scuseria_gaussian_2019} at the B3LYP level with D3(BJ) empirical dispersion corrections and using the 6-311G++(d,p) basis set~\cite{becke_densityfunctional_1993, grimme_consistent_2010, grimme_effect_2011, grimme_effects_2013}. The optimized geometries are depicted in Fig.~\ref{fig:mol_structures} in the SI. Frequency calculations were also performed to acquire predicted quartic centrifugal distortion constants. The structures of the species calculated here follow a similar motif of ring-puckering in the cyclohexene ring as seen elsewhere~\cite{smith_preparation_2020}. One conformer of 1-cyanocyclohexene was found, while cyanocyclohexane and 4-cyanocyclohexene each have chair-equatorial and chair-axial conformers caused by the ring-puckering, which are both expected to be observed in the rotationally cooled molecular jet of each species.

To gain insight in weighing the benefits gained in accuracy for higher levels of theory with the detriment of increased computational resources needed, other commonly used and recommended methods, M06-2X, MP2, and B2PLYPD3~\cite{lee_bayesian_2020}, were also used in combination with the 6-311++G(d,p) and def2TZVP~\cite{marshall_rotational_2017} basis sets to optimize each molecular geometry. The results are discussed in Section~\ref{sec:results} and provided for comparison in Table~\ref{tab:4cn_ax_eq_extra}. 

{Each combination of method and basis set that was tested provides rotational constants with percent errors on the order of $1\,\%$ or better between theory and experiment.} We find that the B3LYP-D3(BJ)/6-311++G(d,p) level of theory provides rotational constants and centrifugal distortion constants that are accurate enough to find the unique spectral pattern of the target species easily, while also using the least amount of computational resources, {reaching optimized geometries on the order of 2-5x faster than the other levels of theory used.} In addition to these geometry optimizations, the rotational constants for all \ce{^{13}C} and \ce{^{15}N} singly-substituted isotopic isomers of {1-cyanocyclohexene and the} 4-cyanocyclohexene conformers were calculated using single-point energy calculations at the B3LYP-D3(BJ)/6-311++G(d,p) level of theory.

\section{Results and Discussion}
\label{sec:results}

\subsection{Frequency Measurement Accuracy and Dynamic Range: OCS}\label{sec:ocs}

\begin{figure}[t!]
    \centering
    \includegraphics[width=\columnwidth]{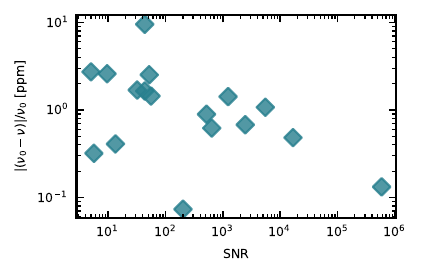}
    \caption{Accuracy in ppm of the $J = 1-0$ rotational transitions of isotopic isomers of \ce{OCS} measured in natural abundance ($\nu$, see Fig.~\ref{fig:ocs_spectra} and Table~\ref{tab:ocs}) in comparison to measurements, $\nu_0$, by the NIST FTMW cavity spectrometer\cite{lovas_pulsed_1987} as a function of SNR.}
    \label{fig:ocs_accuracy}
\end{figure}

We used the simple linear molecule, \ce{OCS}, as a benchmark for assessing the performance of our new compact CP-FTMW spectrometer. The instrumental frequency measurement accuracy can be compared to other spectrometers in the literature~\cite{lovas_pulsed_1987,brown_broadband_2008} and the dynamic range can be demonstrated by detecting signals from isotopic variants of the \ce{OCS} molecule found in natural abundance as low as $0.002\,\%$ (or $20\,\mathrm{ppm}$) of the most naturally occurring isotopologue, yielding the ability to test measurements of signals with a dynamic range of 500000:1. A commercial sample of $0.5\,\%$ \ce{OCS} in Helium (Linde Gas \& Equipment Inc.) was directly attached to the spectrometer and regulated to a backing pressure of $2\,\mathrm{bar}$. The sample was pulsed at a repetition rate of $3\,\mathrm{Hz}$ using a $600\,\mathrm{\upmu s}$ long pulse from the Iota One to open the nozzle, allowing the gas to supersonically expand into the vacuum chamber. A series of ten back-to-back measurements were used per gas introduction yielding an effective measurement repetition rate of 30\,Hz. A $6.5-18.0\,\mathrm{GHz}$ excitation chirp was used; however, for increased SNR, one may use a narrower range of frequency in the excitation chirp. The rest frequencies were determined by fitting Gaussian functions to the lines depicted in Fig.~\ref{fig:ocs_spectra} and are listed in Table~\ref{tab:ocs} in the SI, corresponding to the $J = 1-0$ rotational transition for ten isotopic isomers of \ce{OCS} from the averaged spectrum after 1\,M shots. Comparing these to the values derived from cavity-enhanced FTMW spectroscopy~\cite{lovas_pulsed_1987}, we report small deviations on the order of ${\sim}0.1 - 3\,\mathrm{ppm}$ translating to approximately $1 - 30\,\mathrm{kHz}$ depending on the SNR of each transition considered, as shown in Fig.~\ref{fig:ocs_accuracy}. The ``outlier'' data point at approximately 10\,ppm is due to the fact that we cannot fully resolve the hyperfine splitting present in the $J = 1-0$ transition of \ce{^{17}OCS}, which is ${\sim}130\,\mathrm{kHz}$ and therefore smaller than our measured linewidths in this experiment that have full width at half maxima (FWHM) of the order of $200\,\mathrm{kHz}$. The root mean squared error (RMSE) relative to the NIST cavity data~\cite{lovas_pulsed_1987} for transitions over a SNR of 100 was calculated to be $9.5\,\mathrm{kHz}$, a value we will use to guide our uncertainty discussion in the following section.

\subsection{Benchmark Rotational Spectroscopy of Cyanocyclohexane}

To further benchmark our spectrometer, we recorded the rotational spectrum of cyanocyclohexane, \ce{C6H11-CN}, the fully hydrogenated (saturated) version of benzonitrile (see Fig.~\ref{fig:mol_structures2d} (b)). The spectroscopy of this molecule is well-known and was reported previously using a cavity-enhanced FTMW spectrometer~\cite{durig_microwave_2010}, delivering narrower linewidths and, hence, superior experimental resolution to broadband CP-FTMW spectrometers. The cyanocyclohexane sample was purchased from Sigma Aldrich (cyclohexanecarbonitrile, CAS: 766-05-2, purity $98\,\%$) and used without further purification. The liquid sample was heated to $45^\circ\,\mathrm{C}$ in a modified nozzle reservoir (with details described in Section~\ref{sec:nozzle} in the SI). Argon at $1.7\,\mathrm{bar}$ was flowed over the sample and pulsed at a rate of 3\,Hz, carrying the molecule vapor into the vacuum chamber. After excitation with a $500\,\mathrm{ns}$ long chirped-pulse sweeping $6.5-18.0\,\mathrm{GHz}$, 10 FIDs were recorded probing each gas pulse and averaged to the desired total number of shots before being processed into the resulting FFT spectra.

\begin{figure*}[htb!]
    \centering
    \includegraphics[width=\textwidth]{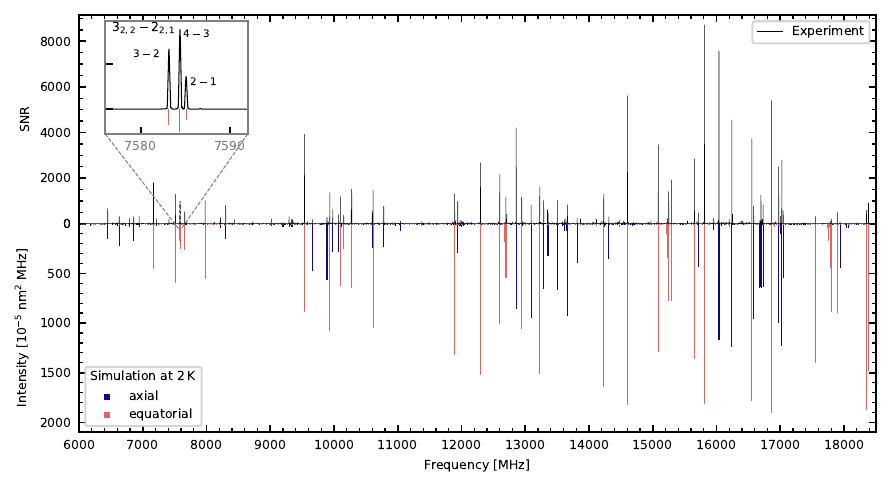}
    \caption{Experimental spectrum of cyanocyclohexane seeded in Ar at 1.5\,M shots. Both conformers are present and their resulting fits are plotted on the inverted $y$-axis (axial -- blue, equatorial -- orange), see Table~\ref{tab:cchexane_ax_eq} for their respective spectroscopic constants. The inset depicts a zoomed version around the $ J_{K_a, K_c} = 3_{2,2} - 2_{2,1}$ transition of the equatorial conformer, with nuclear quadrupole hyperfine splitting caused by the nitrogen atom noted.}
    \label{fig:cyanocyclohexane_spectrum}
\end{figure*}

\begin{figure}[htb!]
    \centering
    \includegraphics[width=\columnwidth]{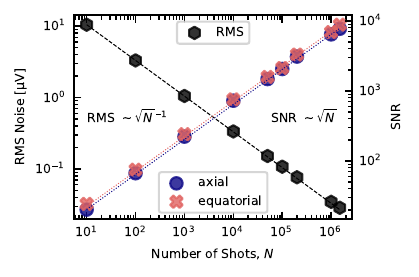}
    \caption{RMS noise (black hexagons) and SNR of two transitions of cyanocyclohexane (blue circles and orange crosses measured on 16038.06 and 15809.49\,MHz belonging to the axial and the equatorial conformer, respectively) as a function of number of shots, {$N$. The dashed and dotted lines represent the proportionality with $\sqrt{N}^{-1}$ and $\sqrt{N}$ for the RMS and SNR, respectively.}}
    \label{fig:rms_snr}
\end{figure}

Figure~\ref{fig:cyanocyclohexane_spectrum} shows the rotational spectrum of cyanocyclohexane averaging 1.5\,M collected FIDs. It is compared to the simulation of the rotational spectra of both conformers (the chair-axial (blue) and chair-equatorial (orange)) at a temperature of 2\,K; both conformers can be identified in the experimental spectrum. We performed the experiment multiple times at a varying number of shots in order to determine the experimental noise floor and SNR on the transitions at 16038.06 and 15809.49\,MHz representing the axial and the equatorial conformer, respectively. The resulting relations are shown in Fig.~\ref{fig:rms_snr} and demonstrate the expected RMS noise decrease with the inverse square root of the number of FIDs averaged (number of shots), $\mathrm{RMS} \sim \sqrt{N}^{-1}$. When reaching of order 1\,M shots, the RMS noise does not decrease significantly with number of shots (which are linear to experiment time).

\subsection{Uncertainty Determination using Gaussian Centers in Rotational Spectra of Cyanocyclohexane}

Fitting Gaussian profiles to lines recorded by CP-FTMW spectrometers is a standard procedure to determine rest frequencies of the molecular signal. Due to spectral noise and finite spectral resolution, the parameters derived from the Gaussian fits (center, amplitude, and width) are each associated with some uncertainty. In least squares fitting to rotational parameters using for example Pickett's \textsc{SPCAT/SPFIT}~\cite{pickett_fitting_1991}, estimates of these uncertainties can be valuable weights that ensure well-defined transition frequencies have the most influence on the resulting spectroscopic constants and that the resulting fits represent the experimental data within experimental uncertainties.

\citet{landman_statistical_1982} derived analytic expressions for the uncertainty of fitted Gaussian parameters in the case of a Gaussian line profile in a white noise environment. For the Gaussian center frequency uncertainty, they derived Eq.~\ref{eqn:landman}, 
\begin{equation}
    \sigma_{x_{0}} = \left( \frac{2}{\pi} \right)^{1/4} \left( \frac{\sigma_i}{A} \right) \sqrt{\sigma \Delta x} \label{eqn:landman}
\end{equation}
which describes the uncertainty, $\sigma_{x_{0}}$, as a function of the width of the fitted Gaussian profile, $\sigma$, and amplitude, $A$, as well as the noise level, $\sigma_i$, and bin width, $\Delta x$, of the spectral data. \citet{golubiatnikov_accuracy_2016} noted that Eq.~\ref{eqn:landman} consistently underestimates the uncertainties in experimental center frequencies as it assumes spectral white noise is the only source of error. Despite this underestimation, Eq.~\ref{eqn:landman} has successfully been used to estimate center frequency uncertainties in rotational fitting procedures\cite{martin-drumel_complete_2015,martin-drumel_millimeter-_2015,cazzoli_precise_2004,cazzoli_sub-doppler_2013,chitarra_pure_2022}. These successes may indicate that Eq.~\ref{eqn:landman} does indeed provide accurate uncertainty estimates when spectral noise is the dominant source of error. To our knowledge, the influences of the Gaussian parameters on the center uncertainty derived by \citet{landman_statistical_1982} have not been verified experimentally. As part of our benchmark analysis of our new CP-FTMW spectrometer, we examine the relationship between center frequency uncertainty and the SNR of the fitted Gaussian. 

Before considering experimental data, Monte Carlo simulations were run to verify that the effects of SNR, peak width, and bin width on the variance of the center of a fitted Gaussian function matched those in Eq.~\ref{eqn:landman}. {Since Gaussian white noise is the only source of error in the idealized Monte Carlo simulations, they are an ideal way to verify the accuracy of the equation derived by \citet{landman_statistical_1982}.} Peaks were simulated under conditions relevant to CP-FTMW spectroscopy: first a Gaussian peak with a known center, FWHM, and amplitude were generated over an array with a known bin width, then Gaussian white noise was randomly generated and added to the simulated Gaussian. For the Gaussian noise, a standard deviation of 1 was chosen for all simulations so that the amplitude of the peak was equivalent to its SNR. Finally, \textsc{scipy}’s \texttt{curve\_fit} was used to fit a Gaussian function to the simulation and the parameters of the fitted function were recorded. This process was repeated 1000 times for each set of center, FWHM, SNR, and bin width, and the variance of the fitted Gaussian functions’ centers were computed for each set of parameters. 

In the Monte Carlo simulations, 60,000 center uncertainties were computed for Gaussian profiles with SNRs ranging from 4 to 1000, FWHMs ranging from 0.3 to 30\,MHz, and bin widths ranging from 0.01 to 0.5\,MHz. These parameter ranges were chosen as a balance between their similarity to CP-FTMW spectra and the computation time required to complete the simulations. The center uncertainties were compared to those predicted by the following equation,

\begin{equation}
    \sigma_{x_0} = \left(\frac{1}{4\pi\ln{2}}\right)^{1/4} \frac{\sqrt{(\text{FWHM})\Delta x}}{\text{SNR}}, \label{eqn:MCpred}
\end{equation}

which is modified from the equation derived by \citet{landman_statistical_1982} to include FWHM and to account for the constant noise level across all trials. Overall, Eq.~\ref{eqn:MCpred} was found to accurately model the uncertainty of the center of the Gaussian function, predicting the uncertainty with a mean absolute percent error of 2.8\,\%. The largest percent errors occurred in cases of poorly defined Gaussian peaks, e.g. when the peak FWHM is smaller than the bin width and when the peak SNR is less than 5.


In our experimental analysis of Gaussian center uncertainties, we focused on determining how a peak’s SNR influences the statistical uncertainty in its fitted center. We chose to focus solely on the SNR because its variability between peaks is significantly greater than the variability of FWHMs and bin widths in experimental spectra recorded by the same instrument. The bin width is insensitive to experimental conditions; with consistent padding of FID data, the bin width of the spectra recorded with the instrument will always be the same. Peak FWHMs are also fairly consistent between spectra recorded with the same supersonic expansion conditions; as discussed in section~\ref{sec:ocs}, we observe FWHMs of approximately 200\,kHz in our experimental spectra. Overlapping transitions produce peaks with the most prominent changes in width, but even in these cases the variability in widths are significantly smaller than the orders of magnitude variability in SNR routinely seen within spectra.

\begin{figure}[t!]
    \centering
    \includegraphics[width=\columnwidth]{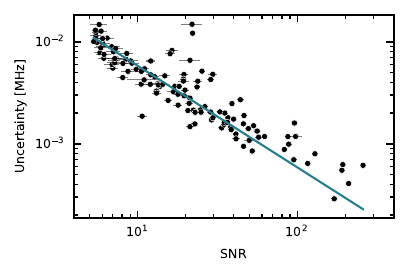}
    \caption{Gaussian center uncertainties as a function of SNR for the peaks observed in 32 experimental rotational spectra of cyanocyclohexane, with a log-log line fit with orthogonal distance regression. Error bars indicate the uncertainty in the SNR of each peak over 32 measurements.}
    \label{fig:gaussian_uncertainty}
\end{figure}

To determine the relationship between SNR and statistical Gaussian center uncertainty in our experimental data, a set of cyanocyclohexane spectra were recorded and the Gaussian fits to peaks in the spectra were analyzed. The cyanocyclohexane spectra were recorded under identical experimental conditions as described above; the liquid sample was heated to $45^\circ\,\mathrm{C}$, argon was pulsed at a backing pressure of $1.7\,\mathrm{bar}$ and a rate of 3\,Hz, 10 FIDs were recorded following each gas pulse, and each spectrum averaged 10k FIDs. To reduce the influence of thermal variations between spectra, the cyanocyclohexane sample was prepared in the instrument a day in advance and preheated to $45^\circ\,\mathrm{C}$ for 20\,hours. 32 spectra were recorded back-to-back, and their FIDs were all processed identically as described above. By preheating the sample and recording all spectra within a few hours of each other, we aimed to minimize variations in {sources of error unrelated to spectral white noise} within the dataset. In total, 122 peaks with SNR above 5 (ranging from 5.3 to 257) were identified and fit with Gaussian profiles in each of the 32 spectra. 


\begin{table*}[ht!]
    \centering
    \caption{{Experimentally determined ground state geometry} spectroscopic constants for axial cyanocyclohexane, {ax. \ce{C6H11-CN}}, and equatorial cyanocyclohexane, {eq. \ce{C6H11-CN}}. All spectroscopic parameters are in MHz except for the permanent electric dipole moment components which are reported in Debye. All {equilibrium geometry} calculations were performed using the 6-311++G(d,p) basis set.}
    \label{tab:cchexane_ax_eq}
    \begin{threeparttable}
    \begin{tabular}{llS[table-format=3.10]S[table-format=3.5]S[table-format=3.5]S[table-format=3.5]lS[table-format=3.8]}
    \multicolumn{8}{c}{\textbf{axial \ce{C6H11-CN}}}\\
    \toprule
        & &\multicolumn{4}{c}{This work} & & \multicolumn{1}{c}{\citet{durig_microwave_2010}} \\
        
        {Parameter} & & {Experimental} & {B3LYP-D3(BJ)} & {M06-2X} & {MP2} & & {Cavity} \\
    \cline{1-1}
    \cline{3-6}
    \cline{8-8}
        &  &  &  &  & & &  \\
        $A$ & & 3004.71428(34) & 3017.8260 & 3011.3467 & 2990.0569 & &  3004.7137(4)   \\
        $B$ & & 1765.19069(24) & 1752.8936 & 1778.7664 & 1770.5469 & & 1765.1907(3)  \\
        $C$ & & 1561.42821(24)  & 1544.0590 & 1573.2529 & 1571.9492 & & 1561.4274(3)  \\
        $\Delta_J \times 10^3$ &&  0.4853(34) & 0.4673 & 0.4642 & 0.4991 & & 0.4878(45) \\
        $\Delta_{JK} \times 10^3$ &&  -1.2271(36) & -1.2442 & -1.1882 & -1.3212 & & -1.2551(243) \\
        $\Delta_K \times 10^3$ && 1.8536(62) & 1.9079 & 1.7769 & 1.9368 & &  1.948(56) \\
        $\delta_J \times 10^3$ && -0.02239(50) & -0.0201 & -0.0197 & -0.0239 & & -0.01916(205) \\
        $\delta_K \times 10^3$ && 0.0491(93) & 0.0123 & -0.0091 & -0.0141 & & 0.087(91) \\
        $\chi_{aa}$ && -1.0721(47) & -1.1844 & -1.1598 & -0.9085 &  & -1.0741(23) \\
        $\chi_{bb}$ &&  2.1290(45)  & 2.2533 & 2.3755 & 1.8308 & & 2.1336(17) \\
        $\chi_{cc}$ &&  & -1.0689  & -1.2156  & -0.9222  \\
        $\chi_{ac}$ &&  & -3.3330  & -3.5088  & -2.7236  \\
        $|\mu_a| / |\mu_b| / |\mu_c|$ && y/n/y\tnote{a} & \text{$3.3/0.0/2.5$} & \text{$3.3/0.0/2.5$} & \text{$3.1/0.0/2.4$}  \\
         &  &  &  &  & \\
        $N_\mathrm{lines}$ && 327 &  & & & &  76 \\
        $\sigma_\mathrm{fit} \times 10^3$ && 11.094 &  &  & \\
        $(J, K_\mathrm{a})_\mathrm{max}$ && {$(13, 7)$} &  & &  & & {$(6,3)$} \\
        \bottomrule
        & \\
        \multicolumn{8}{c}{\textbf{equatorial \ce{C6H11-CN}}}\\
    \toprule
        && \multicolumn{4}{c}{This work} & & \multicolumn{1}{c}{\citet{durig_microwave_2010}} \\
        {Parameter} && {Experimental} & {B3LYP-D3(BJ)} & {M06-2X} & {MP2} & & {Cavity} \\
  \cline{1-1}
    \cline{3-6}
    \cline{8-8}
        &  &  &  &  &  & &\\
        $A$ &&  4239.45449(63) & 4235.0839 & 4256.7810 & 4252.1605 & & 4239.4534(4)  \\
        $B$ &&  1399.18343(19) & 1397.2515 & 1405.6106 & 1393.7657 & & 1399.1834(2)  \\
        $C$ &&  1128.85722(18)  & 1125.9399 & 1133.8621 & 1126.4835 & & 1128.8573(1)  \\
        $\Delta_J \times 10^3$ &&  0.0653(17)  & 0.0614 & 0.0622 & 0.0621 & & 0.0638(17) \\
        $\Delta_{JK} \times 10^3$ &&  0.4192(63) & 0.4091 & 0.4013 & 0.4084 & & 0.432(9) \\
        $\Delta_K \times 10^3$ && 0.688(64) & 0.5577 & 0.5463 & 0.5739 & &  0.535(48) \\
        $\delta_J \times 10^3$ && 0.01053(43) & 0.0106 & 0.0107 & 0.0106 & & 0.01066(87) \\
        $\delta_K \times 10^3$ && 0.354(16) & 0.3304 & 0.3304 & 0.3368 & & 0.308(41) \\
        $\chi_{aa}$& & -3.8651(33)  & -4.1150 & -4.3116 & -3.3346 & & -3.8565(22) \\
        $\chi_{bb}$ &&  2.1185(41)   & 2.2394 & 2.3500 & 1.8111 & & 2.1069(152) \\
        $\chi_{cc}$ &&  & 1.8756 & 1.9617 & 1.5235 & &  \\
        $\chi_{ac}$ & & & 1.4938 & 1.6051 & 1.2380 & & \\
        $|\mu_a| / |\mu_b| / |\mu_c|$ & & y/n/y\tnote{a} & 
        \text{$4.5/0.0/0.7$}  & \text{$4.5/0.0/0.7$}  & \text{$4.3/0.0/0.7$} &  \\
        $\Delta E$\tnote{b} & & & 0.147 & 0.423 & 0.584 & \\
        \\
        $N_\mathrm{lines}$ & & 293 &  & & & &  99 \\
        $\sigma_\mathrm{fit} \times 10^3$ && 9.469 &  &  & & & \\
        $(J, K_\mathrm{a})_\mathrm{max}$ & &{$(13, 6)$} &  &  & & & {$(8, 4)$} \\
        \bottomrule
    \end{tabular}
    \begin{tablenotes}
        \item[a] y or n (yes or no) indicates if a transition type was observed.
        \item[b] Electronic energy (kcal/mol) relative to the axial conformer.
    \end{tablenotes}
    \end{threeparttable}
\end{table*}

In the experimental analysis, the mean SNR and {center frequency variance} of each observed transition were computed using Gaussian profiles fitted to all transitions above a SNR of 5 in each of the 32 identically recorded spectra. {The square roots of the center variances, the standard deviations, were used as estimates of the uncertainty in each experimental measurement.}  Under the assumption that the center uncertainty is proportional to some power of the SNR of the Gaussian profile, the mean SNR and center uncertainty were plotted on a log-log scale and a log-log-linear relation was fit to the data presented in Fig.~\ref{fig:gaussian_uncertainty}. The slope and $y$-intercept of the line were determined with a least squares orthogonal distance regression in order to consider deviation in the SNR values over the 32 spectra. Ultimately, a slope of -1.00(5) and a $y$-intercept of -1.24(5) were determined; the center uncertainties of the Gaussian profiles fitted to peaks observed in our experimental data are then described by

\begin{equation}
    \sigma_{x_0} = \frac{1}{10^{1.24}} \cdot \frac{1}{\text{SNR}}. \label{eqn:fit}
\end{equation}

The inverse relationship between center uncertainty and SNR is consistent with the equation derived by \citet{landman_statistical_1982}. Equation~\ref{eqn:fit} was used in fitting procedures to estimate the statistical uncertainties in the measured transition frequencies.
In our measurements of OCS isotopologue transitions, we calculated a RMSE of $9.5\,\mathrm{kHz}$ relative to NIST cavity data~\cite{lovas_pulsed_1987} for transitions over a SNR of 100. Statistical error due to spectral white noise for these strong peaks was expected to be negligible compared to {alternative sources of error, such as Doppler shifts or timing instability in the electroncies.} Based on the RMSE of the OCS measurements, and following uncertainty estimates of other similar CP-FTMW spectrometers~\cite{crabtree_rotational_2024}, a {base, uniform error of 10\,kHz was assumed for all measured frequencies. The uniform} and spectral noise errors were added in quadrature to estimate the total uncertainty in each measured transition frequency.

In the case of cyanocyclohexane, 327 and 293 transitions for the axial and equatorial conformer, respectively, were recorded and fit using \textsc{SPCAT/SPFIT} in Pickett’s \textsc{CALPGM} suite of programs~\cite{pickett_fitting_1991} to derive their spectroscopic constants by least-squares fitting with a Watson Hamiltonian (A-reduced, $I^r$ representation). Quartic centrifugal distortion and $^{14}$N nuclear electric quadrupole hyperfine coupling constants were included in the fitting procedure. The resulting constants are presented in Table~\ref{tab:cchexane_ax_eq} and compared to the values computed by DFT and the ones measured using a cavity-enhanced FTMW spectrometer~\cite{durig_microwave_2010}. Our experimentally determined constants are in exceptionally good agreement with the previous high-resolution work~\cite{duley_laboratory_2005}, extending the number of recorded transitions by 251 and 194 for the axial and equatorial conformers of cyanocyclohexane, respectively.

\subsection{1-Cyanocyclohexene}

\begin{figure}[t!]
    \centering
    \includegraphics[width=\columnwidth]{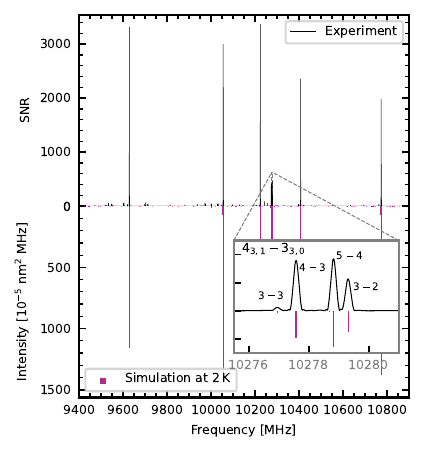}
    \caption{Part of the experimental spectrum of 1-cyanocyclohexene seeded in Ar at 1.5\,M shots in comparison to the fit on the inverted $y$-axis using the spectroscopic constants presented in Table~\ref{tab:1cchexene}. The inset shows a close up on the $ J_{K_a, K_c} = 4_{3,1} - 3_{3,0}$ transition, with hyperfine splitting noted. See Fig.~\ref{fig:1cchexene_full} for the full spectrum.}
    \label{fig:1cchexene}
\end{figure}

The spectrum of 1-cyanocyclohexene, \ce{C6H9}-\ce{1-CN}, was recorded in a similar way to its closely related sibling cyanocyclohexane. 1-cyanocyclohexene (Sigma Aldrich, cyclohexene-1-carbonitrile, CAS: 1855-63-6) was used without further purification (purity $\geq97\,\%$) and heated to $50^\circ\,\mathrm{C}$ in the sample holder. Ar was pulsed at a backing pressure of $1.4\,\mathrm{bar}$ and a $3\,\mathrm{Hz}$ repetition rate. A $300\,\mathrm{ns}$ long chirped-pulse was swept over the $6.5-18\,\mathrm{GHz}$ frequency range and 1.5\,M FIDs were collected. A part of the experimental spectrum is depicted in the upper panel of Fig.~\ref{fig:1cchexene} and compared to the simulation at 2\,K using the resulting spectroscopic constants which are listed in Table~\ref{tab:1cchexene}. The well-resolved $^{14}$N hyperfine splitting of the $J_{K_a, K_c} = 4_{3,1} - 3_{3,0}$ transition is shown in the inset of Fig.~\ref{fig:1cchexene}. While our experimentally derived spectroscopic constants including quartic centrifugal distortion and $^{14}$N nuclear electric quadrupole coupling constants are in good agreement with the theoretically computed values (see Table~\ref{tab:1cchexene}), we were able to improve the spectroscopy when comparing to previous work~\cite{ying-sing_li_microwave_1985}, extending the number of recorded transitions to 184 
and constraining $A$, $B$, and $C$ to lower uncertainties by approximately two orders of magnitude. We also note the presence of chirality for 1-cyanocyclohexene due to the ring-puckering motif within the cyclohexene ring; however, the rotational spectrum of each enantiomer is identical using this technique (see Figs.~\ref{fig:mol_structures2d} and \ref{fig:mol_structures}).

\begin{table*}[htb!]
    \centering
    \caption{{Experimentally determined ground state geometry} spectroscopic constants for 1-cyanocyclohexene, \ce{C6H9}-\ce{1-CN}. All spectroscopic parameters are in MHz. All spectroscopic parameters are in MHz except for the permanent electric dipole moment components which are reported in Debye. All {equilibrium geometry} calculations were performed using the 6-311++G(d,p) basis set.}
    \label{tab:1cchexene}
    \begin{threeparttable}
    \begin{tabular}{llS[table-format=3.10]S[table-format=3.5]S[table-format=3.5]S[table-format=3.5]lS[table-format=3.5]}
    \multicolumn{8}{c}{\textbf{equatorial \ce{C6H9}-\ce{1-CN}}}\\
    \toprule
        & & \multicolumn{4}{c}{This work} & & \multicolumn{1}{c}{\citet{ying-sing_li_microwave_1985}} \\
        
        {Parameter} && {Experimental} & {B3LYP-D3(BJ)} & {M06-2X} & {MP2} & & {Absorption} \\
  \cline{1-1}
    \cline{3-6}
    \cline{8-8}
        &  &  &  &  &  & &\\
        $A$ & & 4565.81114(87) & 4574.0613 & 4586.8101 & 4568.5288 & & 4565.98(46)  \\
        $B$ &  & 1423.45942(44) & 1424.3204 & 1429.7802 & 1416.4996 & & 1423.44(1) \\
        $C$ &  & 1136.17833(39) & 1135.8762 & 1141.2691 & 1132.9468 & & 1136.17(1) \\
        $\Delta_J \times 10^3$ & & 0.0471(39) & 0.0463 & 0.0467 & 0.0463   \\
        $\Delta_{JK} \times 10^3$ & & 0.646(14) & 0.6039 & 0.6104 & 0.6335   \\
        $\Delta_K \times 10^3$ & &  & 0.3022 & 0.2876 & 0.2680   \\
        $\delta_J \times 10^3$ & & 0.0122(13) & 0.0092 & 0.0093 & 0.0091   \\
        $\delta_K \times 10^3$ & & 0.384(79) & 0.3545 & 0.3602 & 0.3681  \\
        $\chi_{aa}$ & & -4.1923(48)  & -4.4321 & -4.6841 & -3.6106 & & -4.2 \\
        $\chi_{bb}$ & & 2.2380(68) & 2.4002 & 2.5099 & 1.8334 & &  \\
        $\chi_{cc}$ & &  & 2.0319 & 2.1742 & 1.7772 & & \\
        $\chi_{ab}$ & &  & 0.1529 & 0.1362 & 0.0790 & & \\
        $\chi_{ac}$ & &  & -0.1126 & -0.1330 & -0.1203 & & \\
        $\chi_{bc}$ & &  & -0.0666 & -0.0616 & -0.0232 & & \\
        $|\mu_a| / |\mu_b| / |\mu_c|$&& y/y/y\tnote{a} & $\text{$5.0/0.2/0.1$}$ & $\text{$5.0/0.2/0.1$}$ & $\text{$4.6/0.1/0.1$}$ &  \\
        \\
        $N_\mathrm{lines}$ & & 184 &  &  &  & & 20 \\
        $\sigma_\mathrm{fit} \times 10^3$ &  & 11.609 &  &  & &  \\
        $(J, K_\mathrm{a})_\mathrm{max}$ & & {$(12, 6)$} &  &  &  & & {$(10,2)$}  \\
        \bottomrule
    \end{tabular}
    \begin{tablenotes}
        \item[a] y or n (yes or no) indicates if a transition type was observed.
    \end{tablenotes}
    \end{threeparttable}
\end{table*}

\subsection{4-Cyanocyclohexene}

\begin{figure*}[htb!]
    \centering
    \includegraphics[width=\textwidth]{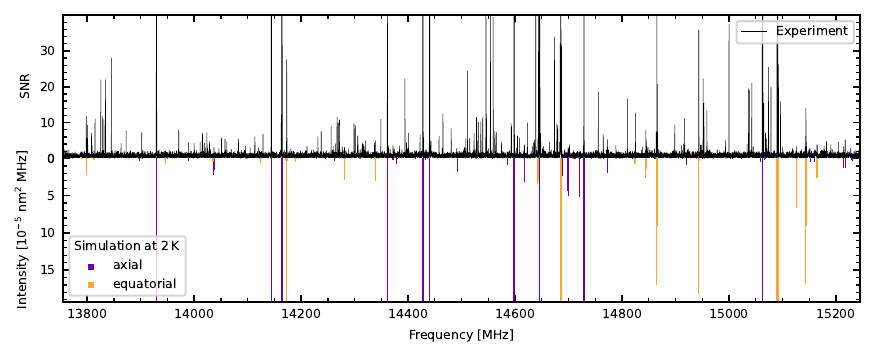}
    \caption{Part of the experimental spectrum of 4-cyanocyclohexene seeded in Ar at 1.5\,M shots in comparison to the fits of the axial (violet) and equatorial (yellow) conformers on the inverted $y$-axis using the spectroscopic constants presented in Table~\ref{tab:4cchexene_ax_eq}. {The majority of remaining unassigned lines below} a SNR of 35 demonstrates the dynamic range and sensitivity of the instrument, as all \ce{^{13}C} and the \ce{^{15}N} {singly-substituted isotopic isomers of axial and equatorial 4-cyanocyclohexene are observable.} See Fig.~\ref{fig:4cchexene_full} for the full spectrum.}
    \label{fig:4cchexene}
\end{figure*}

\begin{figure*}[htb!]
    \centering
    \includegraphics[width=\textwidth]{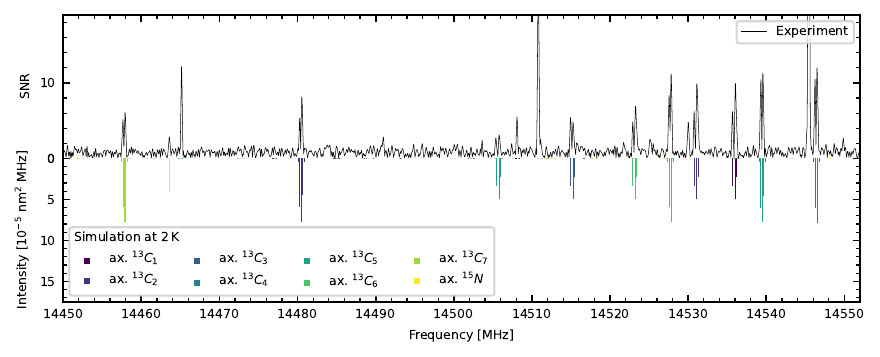}
    \caption{Similar to Fig.~\ref{fig:4cchexene} showing {a sample of assigned} \ce{^{13}C} and the \ce{^{15}N} substituted isotopic isomers of the axial conformer of 4-cyanocyclohexene, {with nuclear hyperfine splitting observable.} The stick spectra are weighed to $1.1\,\%$ and $0.4\,\%$ in accordance with the natural abundance of \ce{^{13}C} and \ce{^{15}N}, respectively. Their spectroscopic constants are listed in Table~\ref{tab:4cchexene_eq_ax_isos}.}
    \label{fig:4cchexene_13C}
\end{figure*}

After successfully benchmarking our spectrometer, we turned to the last member of the family of commercially available cyano-substituted cyclohexenes and cyclohexanes. In contrast to its siblings, the chiral 4-cyanocyclohexene, {\ce{C6H9}-\ce{4-CN}}, has not been studied previously by rotational spectroscopy. The 4-cyanocyclohexene sample was purchased from TCI America, Inc. (4-cyano-1-cyclohexene, CAS: 100-45-8) and used without further purification (purity $>98\,\%$). The liquid sample was loaded into the modified nozzle sample reservoir and heated to $55^\circ\,\mathrm{C}$. Ar with an absolute backing pressure of $1.7\,\mathrm{bar}$ was used as the carrier gas in the pulsed jet sample introduction. A $600\,\mathrm{ns}$ chirped-pulse spanning $6.5-18\,\mathrm{GHz}$ was used to polarize the sample, and 10 FIDs were recorded and averaged per sample introduction gas pulse until a total average of 1.5\,M FIDs was obtained. A part of the fast Fourier transformed spectrum is depicted in Fig.~\ref{fig:4cchexene}. As expected from the quantum chemical calculations, the spectra of the axial and equatorial conformers are observed (see Fig.~\ref{fig:mol_structures2d}). Experimental fits to both conformers are determined and listed in Table~\ref{tab:4cchexene_ax_eq}.

{
\subsection{Structural Analysis}}

\begin{figure*}[htb!]
    \centering
    \includegraphics[width=\textwidth]{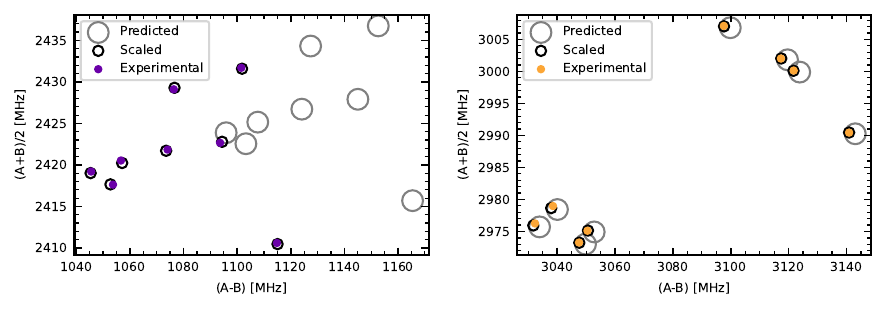}
    \caption{Representation of rotational constants determined for all \ce{^{13}C} and \ce{^{15}N} substituted isotopic isomers of 4-cyanocyclohexene in its axial (left, violet) and equatorial (right, yellow) conformation as presented analogously to Holdren,~PhD~thesis~2022~\cite{holdren_quantitative_2022}. We compare the predicted constants (grey circles) calculated at the B3LYP-D3(BJ)/6-311++G(d,p) level of theory to the experimentally scaled (black circles) and experimentally determined (colored dots) values. {The size of the circles is a qualitative representation of increasing accuracy from the predicted values by scaling the rotational constants which more closely match the experimentally determined constants} (see Table~\ref{tab:4cchexene_eq_ax_isos}).}
    \label{fig:4cchexene_isos}
\end{figure*}

{The spectra of all singly-substituted \ce{^{13}C} and \ce{^{15}N} isotopic isomers at approximately 1.1\,\% and 0.4\,\% intensity of the parent isotopologue were observed for 1-cyanocyclohexene and both conformers of 4-cyanocyclohexene due to both the high sensitivity of the instrument and the time spent averaging (see Figs.~\ref{fig:4cchexene} and \ref{fig:4cchexene_13C} for the 4-cyanocyclohexene isotopologues). The calculated rotational constants of each isotopic isomer were scaled using the error between the calculated constants and the experimentally determined constants of the parent isotopologue (see Table~\ref{tab:4cchexene_ax_eq_isos_calc} and Fig.~\ref{fig:4cchexene_isos}). This approach provides theoretical rotational constants that are more accurate~\citep{tikhonov_scaling_2024}, with percent errors compared to experimentally determined constants on the order of $10-100\times$ lower. Using these scaled rotational constants and retaining the nuclear hyperfine coupling constants and centrifugal distortion constants of the respective parent isotopolgoues, the eight isotopic isomers of each species were experimentally determined. Their resulting rotational constants are listed in Table~\ref{tab:4cchexene_eq_ax_isos}. 
Although the rotational constants are quite similar between all singly-substituted isotopic isomers of a given species, the low uncertainty in their experimental fits and comparison of percent error to each other isotopic isomer species' fit is proof of their discernibility as shown in Fig.~\ref{fig:4cchexene_isos} by the relative separation of each species without overlap. The strong improvement between theory and experiment after scaling theoretical constants is demonstrated.} 

{The shifts in the principal rotational constants of each isotopic isomer relative to the parent isotopologues provide insight into the exact positions of the heavy atoms in 1-cyanocyclohexene and each 4-cyanocyclohexene conformer. A Kraitchman analysis was performed with Kisiel's \textsc{KRA} program to obtain the Cartesian coordinates of every carbon and nitrogen atom relative to the center of mass of the respective molecule using the experimental rotational constants of each isotopic isomer~\cite{kraitchman_determination_1953,kisiel_prospe_nodate}. Uncertainties in the positions of each carbon and nitrogen atom were determined by propagating the uncertainties in the rotational constants and are corrected with the Costain rule~\cite{costain_further_1966}. Experimentally determined coordinates of all heavy atoms are presented in Table~\ref{tab:ax_eq_kraitchman_coords}. Kisiel's \textsc{EVAL} program was then used to determine bond lengths and bond angles between these heavy atoms which are shown in Table~\ref{tab:ax-eq_structure_params}, and these structural parameters for the heavy atom substitution structure are compared to the calculated equilibrium geometry.}

{The axial 4-cyanocyclohexene conformer is comparable to theoretical values and is determined with low uncertainties. The equatorial 4-cyanocyclohexene conformer and 1-cyanocyclohexene yield higher uncertainties in atomic coordinates and bond angles, especially those related to the \ce{CN} group. These uncertainties can largely be attributed to several atoms in each molecule falling closely upon principal axes in the molecular coordinate system. Due to the $r^2$ dependence of moments of inertia, when close to an axis, small changes in the position of an atom result in even smaller changes in moments of inertia\cite{costain_determination_1958}. The small shifts in these atom positions could also have a sign difference across the principal axis which compounds the uncertainty (see Table~\ref{tab:ax_eq_kraitchman_coords}). More precise rotational constants in relevant isotopologues are then required to calculate accurate substitution coordinates. In both molecules, the cyano group is aligned along the a-axis which makes it challenging to determine bond angles associated with the cyano group and the cyano-substituted carbon atom in the ring. Additionally, due to the near-planarity of the two molecules, multiple atoms have small b-axis coordinates, which result in more uncertain bond lengths around the rings. These issues are not uncommon for semi-planar molecules, as seen in ethynylcyclohexane~\cite{vogt_semiexperimental_2018} and in other nearly planar structures~\cite{salvitti_structure_2024,lei_laboratory_2023,neeman_gas-phase_2025}. Overall, the spectral fits for all singly-substituted \ce{^{13}C} and \ce{^{15}N} isotopic isomers of 1-cyanocyclohexene and for each conformer of 4-cyanocyclohexene, alongside the strong comparison to theoretical structural parameters, demonstrate the ability of the spectrometer to determine absolute structural information with high accuracy and precision.}

\begin{figure}[b!]
    \centering
    \includegraphics[width=\columnwidth]{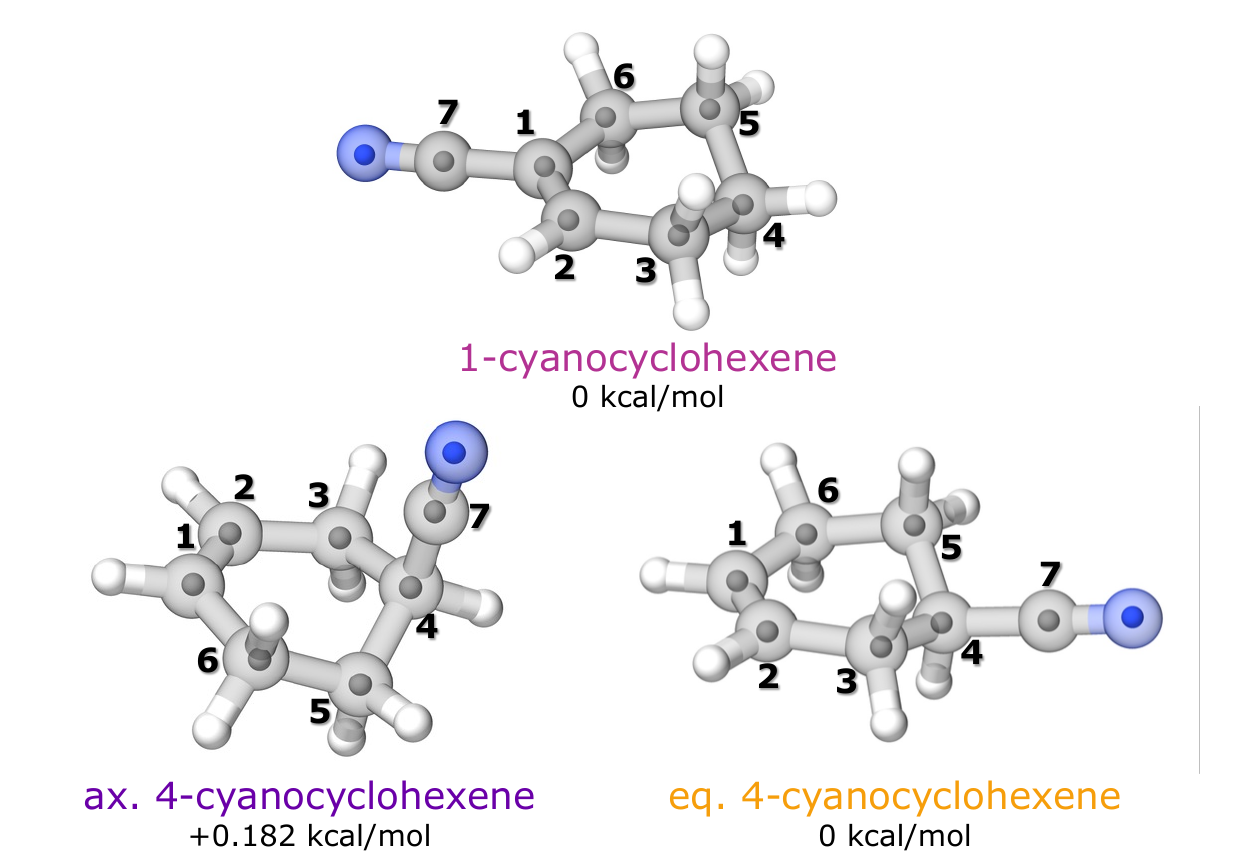}
    \caption{B3LYP-D3(BJ)/6-311++G(d,p) optimized {equilibrium geometries of 1-cyanocyclohexene (top) and} 4-cyanocyclohexene in its axial (left) and equatorial (right) conformation displayed as transparent spheres. Carbon atoms are numbered, with the \ce{C=C} double bond between $C_\mathrm{1}$ and $C_\mathrm{2}$, and relative energy between conformers noted. Smaller, solid spheres represent the experimentally determined heavy atom positions found through a Kraitchman analysis using spectroscopic fits of each singly-substituted isotopic isomer found in natural abundance.}
    \label{fig:14cchexene_labeling}
\end{figure}

{The substitution structures for 1-cyanocyclohexene and 4-cyanocyclohexene (see Fig.~\ref{fig:14cchexene_labeling}) provide an opportunity to compare how the hybridization of the cyano-substituted carbon influences the overall structure. The \ce{C#N} triple bond in 1-cyanocyclohexene can conjugate with the neighboring double bond, potentially resulting in electrons that are delocalized over the four atoms. In the 1-cyanocyclohexene substitution structure, the \ce{C#N} bond is slightly lengthened and the C-\ce{CN} bond between the cyano group and the $sp^2$ carbon in the ring is slightly shortened compared to the 4-cyanocyclohexene substitution structures. The C-CN bond shows the most prominent change between the two isomers, and at 1.4489(35)\,Å in 1-cyanocyclohexene, the bond length is nearly the average of typical \ce{C-C} single bonds (1.53\,Å) and \ce{C=C} double bonds (1.34\,Å)~\cite{lide_survey_1962}. These changes are consistent with a small decrease in the \ce{C#N} bond order and an increase in the C-CN bond order in 1-cyanocyclohexene. Despite these trends, the large uncertainties on the experimental bond lengths make it difficult to definitively determine the degree of delocalization.}\\


\begin{table*}[htb!]
    \centering
    \caption{{Experimentally determined ground state geometry} spectroscopic constants for axial 4-cyanocyclohexene, {ax. \ce{C6H9}-\ce{4-CN}}, and equatorial 4-cyanocyclohexene, {eq. \ce{C6H9}-\ce{4-CN}}. All spectroscopic parameters are in MHz. All spectroscopic parameters are in MHz except for the permanent electric dipole moment components which are reported in Debye. All {equilibrium geometry} calculations were performed using the 6-311++G(d,p) basis set.}
    \label{tab:4cchexene_ax_eq}
    \begin{threeparttable}
    \begin{tabular}{llS[table-format=3.10]S[table-format=3.5]S[table-format=3.5]S[table-format=3.5]}
    \multicolumn{6}{c}{\textbf{axial \ce{C6H9}-\ce{4-CN}}}\\
    \toprule
        \multicolumn{6}{c}{This work}  \\
        {Parameter} && {Experimental} & {B3LYP-D3(BJ)} & {M06-2X} & {MP2} {} \\
  \cline{1-1}
    \cline{3-6}
        &  &  &  & &\\
        $A$ && 2982.81120(50) & 3013.3694 & 2991.6132 & 2954.6649   \\
        $B$ && 1899.16757(39) & 1878.7227 & 1912.7221 & 1909.7925  \\
        $C$ && 1696.22518(42) & 1668.6924 & 1707.7674 & 1714.0239 \\
        $\Delta_J \times 10^3$ && 0.7295(62) & 0.6862 & 0.6954 & 0.7089 \\
        $\Delta_{JK} \times 10^3$ && -1.831(11) & -2.4655 & -1.8527 & -1.8426  \\
        $\Delta_K \times 10^3$ && 2.316(15) & 2.4655 & 2.3121 & 2.2547 \\
        $\delta_J \times 10^3$ && -0.0554(31) & -0.0533 & -0.0544 & -0.0575 \\
        $\delta_K \times 10^3$ &&  & 0.1259 & 0.1203 & 0.1267 \\
        $\chi_{aa}$ && -0.9657(56) & -1.0884 & -1.0530 & -0.8219 \\
        $\chi_{bb}$ && 1.9983(51) & 2.1480 & 2.2311 & 1.7018 \\
        $\chi_{cc}$ &&  & -1.0597 & -1.1781 & -0.8799 \\
        $\chi_{ab}$ &&  & 0.6108 & 0.7026 & 0.5876 \\
        $\chi_{ac}$ &&  & -3.2837 & -3.4448 & -2.6633 \\
        $\chi_{bc}$ &&  & 0.6155 & 0.7275 & 0.5914 \\
        $|\mu_a| / |\mu_b| / |\mu_c|$ && y/y/y\tnote{a} & \text{$3.1/0.2/2.7$} & \text{$3.0/0.3/2.7$}  & \text{$2.9/0.3/2.6$}  \\
        $\Delta E$\tnote{b} & &  & 0.182 & 0.001 & 0.439  \\
        \\
        $N_\mathrm{lines}$ && 151 &  &  &   \\
        $\sigma_\mathrm{fit} \times 10^3$ && 10.917 &  &  &   \\
        $(J, K_\mathrm{a})_\mathrm{max}$ && {$(10, 6)$} &  &  &    \\
        \bottomrule
        & \\
    \multicolumn{6}{c}{\textbf{equatorial \ce{C6H9}-\ce{4-CN}}}\\
    \toprule
         \multicolumn{6}{c}{This work}  \\
        {Parameter} && {Experimental} & {B3LYP-D3(BJ)} & {M06-2X} & {MP2}\\
  \cline{1-1}
    \cline{3-6}
        &  & & &  &\\
        $A$ && 4561.31154(92) & 4562.1851 & 4581.6442 & 4564.7422  \\
        $B$ && 1460.14323(49) & 1458.8389 & 1466.5777 & 1453.8192  \\
        $C$ && 1160.76780(41) & 1158.8045 & 1165.7604 & 1157.5948 \\
        $\Delta_J \times 10^3$ && 0.0639(41) & 0.0594 & 0.0598 & 0.0584 \\
        $\Delta_{JK} \times 10^3$ && 0.663(15) & 0.6176 & 0.6230 & 0.6402  \\
        $\Delta_K \times 10^3$ && 0.241(95) & 0.2568 & 0.2324 & 0.2092 \\
        $\delta_J \times 10^3$ && 0.0107(25) & 0.0119 & 0.0120 & 0.0117 \\
        $\delta_K \times 10^3$ && 0.458(80) & 0.4206 & 0.4245 & 0.4293 \\
        $\chi_{aa}$ &&  -4.0066(41) & -4.2591 & -4.4754 & -3.4765 \\
        $\chi_{bb}$ && 2.1228(54) & 2.2542 & 2.3628 & 1.8149 \\
        $\chi_{cc}$ &&  & 2.0048 & 2.1126 & 1.6616 \\
        $\chi_{ab}$ &&  & -0.0291 & -0.0339 & -0.0341 \\
        $\chi_{ac}$ &&  & -1.2007 & -1.2708 & -0.9380 \\
        $|\mu_a| / |\mu_b| / |\mu_c|$ && y/y/y\tnote{a} & \text{$4.2/0.2/0.5$}  & \text{$4.2/0.2/0.5$}  & \text{$4.0/0.2/0.5$} \\
        \\
        $N_\mathrm{lines}$ && 213 &  &  &  \\
        $\sigma_\mathrm{fit} \times 10^3$ && 9.836 &  &  &  \\
        $(J, K_\mathrm{a})_\mathrm{max}$ && {$(9, 6)$} &  &  &   \\
        \bottomrule
    \end{tabular}
        \begin{tablenotes}
        \item[a] y or n (yes or no) indicates if a transition type was observed.
        \item[b] Electronic energy (kcal/mol) relative to the equatorial conformer.
    \end{tablenotes}
    \end{threeparttable}
\end{table*}

\begin{table*}[htb!]
    \centering
    \caption{{Experimentally determined ground state geometry} spectroscopic constants for the singly-substituted isotopic isomers of 1-cyanocyclohexene (\ce{C6H9}-\ce{1-CN}), axial 4-cyanocyclohexene (ax. \ce{C6H9}-\ce{4-CN}), and equatorial 4-cyanocyclohexene (eq. \ce{C6H9}-\ce{4-CN}), found in natural abundance in the recorded spectrum of the parent species. Centrifugal distortion constants and nuclear quadrupole coupling constants are fixed to those of the parent isotopologues. All spectroscopic parameters are in MHz.}
    \label{tab:4cchexene_eq_ax_isos}
    \begin{threeparttable}
    \begin{tabular}{lcS[table-format=3.10]S[table-format=3.10]S[table-format=3.10]S[table-format=3.10]S[table-format=3.10]}

        \multicolumn{6}{c}{\textbf{\ce{C6H9}-\ce{1-CN}}}\\
        \toprule
        {Parameter} && 
        {${}^{13}C_\mathrm{1}$} & 
        {${}^{13}C_\mathrm{2}$} & 
        {${}^{13}C_\mathrm{3}$} & 
        {${}^{13}C_\mathrm{4}$} \\
  \cline{1-1}
    \cline{3-6}
        & & &  &  & \\
        $A$ && 4565.938(27) & 4506.123(50) & 4498.158(51) & 4560.493(44) \\
        $B$ && 1422.02290(42) & 1423.44160(76) & 1413.88259(85) & 1403.85587(65) \\
        $C$ && 1135.26870(38) & 1132.48417(66)  & 1125.90882(66) & 1124.00095(57) \\
        $N_\mathrm{lines}$ && 46 & 45 & 35 & 44 \\
        $\sigma_\mathrm{fit} \times 10^3$ && 7.227 & 11.081 & 11.299 & 11.625 \\
        & & &  &  & \\
        {Parameter} && 
        {${}^{13}C_\mathrm{5}$} & 
        {${}^{13}C_\mathrm{6}$} & 
        {${}^{13}C_\mathrm{7}$} & 
        {${}^{15}N$} \\
  \cline{1-1}
    \cline{3-6}
        & & &  &  & \\
        $A$ && 4504.053(45) & 4497.163(43) & 4565.681(43) & 4565.682(55) \\
        $B$ && 1413.62969(66) & 1423.45174(73) & 1406.92159(71) & 1383.84455(79) \\
        $C$ && 1126.54472(52) & 1131.90240(68) & 1125.61293(61)  & 1110.79791(70) \\ 
        $N_\mathrm{lines}$ && 32 & 49 & 50 & 28 \\
        $\sigma_\mathrm{fit} \times 10^3$ && 7.698 & 13.316 & 9.149 & 12.234 \\
        \bottomrule
        & \\
        \multicolumn{6}{c}{\textbf{axial \ce{C6H9}-\ce{4-CN}}}\\
        \toprule
        {Parameter} && 
        {${}^{13}C_\mathrm{1}$} & 
        {${}^{13}C_\mathrm{2}$} & 
        {${}^{13}C_\mathrm{3}$} & 
        {${}^{13}C_\mathrm{4}$} \\
  \cline{1-1}
    \cline{3-6}
        & & &  &  & \\
        $A$ && 2969.4859(19) & 2958.8750(14) & 2941.9886(14) & 2967.2905(14) \\
        $B$ && 1875.84622(77) & 1884.81370(81) & 1896.42339(97) & 1890.9266(11) \\
        $C$ && 1681.5135(11) & 1679.78822(86) & 1683.99794(83) & 1694.03366(87) \\
        $N_\mathrm{lines}$ && 25 & 32 & 35 & 28 \\
        $\sigma_\mathrm{fit} \times 10^3$ && 11.922 & 9.900 & 11.316 & 10.439 \\
        & & &  &  & \\
        {Parameter} && 
        {${}^{13}C_\mathrm{5}$} & 
        {${}^{13}C_\mathrm{6}$} & 
        {${}^{13}C_\mathrm{7}$} & 
        {${}^{15}N$} \\
  \cline{1-1}
    \cline{3-6}
        & & &  &  & \\
        $A$ && 2944.4751(15) & 2948.8661(24) & 2982.4702(51) & 2968.0286(32) \\
        $B$ && 1890.7753(11) & 1892.2022(11) & 1880.9279(11) & 1853.1794(10) \\
        $C$ && 1689.63843(84) & 1680.2110(11) & 1681.7374(11) & 1663.92606(70) \\ 
        $N_\mathrm{lines}$ && 31 & 24 & 18 & 15 \\
        $\sigma_\mathrm{fit} \times 10^3$ && 11.231 & 11.320 & 11.138 & 11.337 \\
        \bottomrule 
        & \\
        \multicolumn{6}{c}{\textbf{equatorial \ce{C6H9}-\ce{4-CN}}}\\
    \toprule
        {Parameter} && 
        {${}^{13}C_\mathrm{1}$} & 
        {${}^{13}C_\mathrm{2}$} & 
        {${}^{13}C_\mathrm{3}$} & 
        {${}^{13}C_\mathrm{4}$} \\
  \cline{1-1}
    \cline{3-6}
        & & &  &  & \\
        $A$ && 4561.147(55) & 4500.287(46)  & 4492.342(84) & 4556.172(76) \\
        $B$ && 1439.25877(81) & 1449.89533(75) & 1460.1130(18) & 1458.3083(13) \\
        $C$ && 1147.52691(75) & 1150.36410(80) & 1156.2879(15) & 1159.9800(13) \\
        $N_\mathrm{lines}$ && 42 & 42 & 26 & 38 \\
        $\sigma_\mathrm{fit} \times 10^3$ && 12.047 & 10.266 & 11.203 & 15.256 \\
        & & &  &  & \\
        {Parameter} && 
        {${}^{13}C_\mathrm{5}$} & 
        {${}^{13}C_\mathrm{6}$} & 
        {${}^{13}C_\mathrm{7}$} & 
        {${}^{15}N$} \\
  \cline{1-1}
    \cline{3-6}
        & & &  &  & \\
        $A$ && 4498.36(12) & 4497.014(63) & 4560.68(10) & 4560.934(63) \\
        $B$ && 1459.7184(23) & 1449.5108(11) & 1443.3442(18) & 1419.98295(89) \\
        $C$ && 1156.8878(20) & 1149.9271(11) & 1150.1646(15) & 1135.26053(82) \\ 
        $N_\mathrm{lines}$ && 23 & 32 & 29 & 16 \\
        $\sigma_\mathrm{fit} \times 10^3$ && 14.756 & 12.914 & 14.945 & 10.894 \\
        \bottomrule
    \end{tabular}
    \end{threeparttable}
\end{table*}

\begin{table*}[htb!]
    \caption{{Experimentally determined structural parameters of 1-cyanocyclohexene (\ce{C6H9}-\ce{1-CN}), axial 4-cyanocyclohexene (ax. \ce{C6H9}-\ce{4-CN}), and equatorial 4-cyanocyclohexene (eq. \ce{C6H9}-\ce{4-CN}), through Kraitchman analysis compared to calculated equilibrium geometry values at the B3LYP-D3(BJ)/6-311G++(d,p) level of theory. Bond lengths are in Angstroms (\r{A}) and bond angles in degrees (\textdegree).}}
    \label{tab:ax-eq_structure_params}
    \begin{threeparttable}
    \begin{tabular}{ccS[table-format=3.5]S[table-format=3.3]cccS[table-format=3.5]S[table-format=3.3]cS[table-format=3.5]S[table-format=3.3]}
     & & \multicolumn{2}{c}{\textbf{\ce{C6H9}-\ce{1-CN}}} &&&& \multicolumn{2}{c}{\textbf{axial \ce{C6H9}-\ce{4-CN}}} && \multicolumn{2}{c}{\textbf{equatorial \ce{C6H9}-\ce{4-CN}}}\\
    \toprule
        {Parameter} && {Exp.\tnote{a}} & {Theo.} && {Parameter} && {Exp.\tnote{a}} & {Theo.} && {Exp.\tnote{a}} & {Theo.} \\
  \cline{1-1} \cline{3-4} \cline{6-6} \cline{8-9} \cline{11-12}

        && &  \\
       $r(C_\mathrm{1}-C_\mathrm{2})$ && 1.355(28) & 1.342  && $r(C_\mathrm{1}-C_\mathrm{2})$ && 1.3242(84) & 1.333 && 1.370(26) & 1.333  \\
       $r(C_\mathrm{2}-C_\mathrm{3})$ && 1.551(21) & 1.500  && $r(C_\mathrm{2}-C_\mathrm{3})$ && 1.5141(44) & 1.506 && 1.564(27) & 1.505  \\
       $r(C_\mathrm{3}-C_\mathrm{4})$ && 1.522(19) & 1.533  && $r(C_\mathrm{3}-C_\mathrm{4})$ && 1.5350(60) & 1.545 && 1.492(20) & 1.543  \\
       $r(C_\mathrm{4}-C_\mathrm{5})$ && 1.530(17) & 1.531  && $r(C_\mathrm{4}-C_\mathrm{5})$ && 1.5233(65) & 1.544 && 1.520(18) & 1.543  \\
       $r(C_\mathrm{5}-C_\mathrm{6})$ && 1.592(25) & 1.532  && $r(C_\mathrm{5}-C_\mathrm{6})$ && 1.5167(66) & 1.531 && 1.573(23) & 1.532  \\
       $r(C_\mathrm{6}-C_\mathrm{1})$ && 1.436(29) & 1.517  && $r(C_\mathrm{6}-C_\mathrm{1})$ && 1.5226(77) & 1.505 && 1.474(26) & 1.506  \\
       $r(C_\mathrm{1}-C_\mathrm{7})$ && 1.4489(35) & 1.430  && $r(C_\mathrm{4}-C_\mathrm{7})$ && 1.4807(92) & 1.465 && 1.4772(39) & 1.461  \\
       $r(C_\mathrm{7}-N)$ && 1.1587(14) & 1.156 && $r(C_\mathrm{7}-N)$ && 1.1512(93) & 1.153 && 1.1575(44) & 1.154  \\
       & \\ 
       $\angle(C_\mathrm{1} C_\mathrm{2}C_\mathrm{3})$ && 119.5(10) & 123.28 && $\angle(C_\mathrm{1} C_\mathrm{2}C_\mathrm{3})$ && 123.79(20) & 123.75 && 122.69(63) & 123.76  \\
       $\angle(C_\mathrm{2} C_\mathrm{3}C_\mathrm{4})$ && 112.64(50) & 112.58 && $\angle(C_\mathrm{2} C_\mathrm{3}C_\mathrm{4})$ && 111.42(18) & 112.19 && 108.9(10) & 111.21  \\
       $\angle(C_\mathrm{3} C_\mathrm{4}C_\mathrm{5})$ && 110.75(47) & 110.69 && $\angle(C_\mathrm{3} C_\mathrm{4}C_\mathrm{5})$ && 111.05(29) & 110.19 && 113.76(94) & 110.47  \\
       $\angle(C_\mathrm{4} C_\mathrm{5}C_\mathrm{6})$ && 111.14(51) & 110.92 && $\angle(C_\mathrm{4} C_\mathrm{5}C_\mathrm{6})$ && 111.06(17) & 111.20 && 107.96(80) & 110.07  \\
       $\angle(C_\mathrm{5} C_\mathrm{6}C_\mathrm{1})$ && 109.7(13) & 111.56 && $\angle(C_\mathrm{5} C_\mathrm{6}C_\mathrm{1})$ && 111.65(27) & 112.05 && 112.80(75) & 112.31  \\
       $\angle(C_\mathrm{6} C_\mathrm{1}C_\mathrm{2})$ && 128.9(12) & 123.33 && $\angle(C_\mathrm{6} C_\mathrm{1}C_\mathrm{2})$ && 123.20(22) & 123.62 && 123.42(19) & 123.56  \\
       $\angle(C_\mathrm{2} C_\mathrm{1}C_\mathrm{7})$ && 114.4(22) & 119.93 && $\angle(C_\mathrm{3} C_\mathrm{4}C_\mathrm{7})$ && 109.32(92) & 110.89 && 108.9(35) & 111.00  \\
       $\angle(C_\mathrm{6} C_\mathrm{1}C_\mathrm{7})$ && 116.7(23) & 116.74  && $\angle(C_\mathrm{5} C_\mathrm{4}C_\mathrm{7})$ && 110.8(11) & 110.44  && 108.8(32) & 111.46  \\
       $\angle(C_\mathrm{4} C_\mathrm{7}N)$ && 177.4(36) & 177.93 && $\angle(C_\mathrm{4} C_\mathrm{7}N)$         && 178.7(25) & 179.38 && 178.7(17) & 179.07\\
        \bottomrule
    \end{tabular}
    \begin{tablenotes}
        \item[a] Value in parentheses is the standard error in least significant figure.
    \end{tablenotes}
    \end{threeparttable}
\end{table*}

\section{Astronomical Analysis and Astrochemical Implications}

\subsection{Results of Astronomical Search}
Rotational spectra of the five cyanocyclohexane and cyanocyclohexene structures were simulated under conditions of the cold, dark molecular cloud TMC-1 ($5-10\,\mathrm{K}$, $5.8\,\mathrm{km/s}$) and compared to radio astronomical observations of this astronomical source conducted as part of the GOTHAM (GBT Observations of TMC-1: Hunting Aromatic Molecules) large program~\cite{mcguire_early_2020}. The observations used are taken by the 100\,m Robert C. Byrd Green Bank Telescope (GBT) pointing at the TMC-1 cyanopolyyne peak (CP) at right ascension 04$^\textrm{h}$41$^\textrm{m}$42$\overset{\scriptsize\raise0.1em\hbox{s}}{.}$50, declination +25$^{\circ}$41$^{\prime}$26$\overset{''}{.}$8 (J2000 equinox). The high-sensitivity, high-spectral resolution GOTHAM survey spectra cover receiver bands from 3.9 to 36.4\,GHz
, totaling approximately 29\,GHz of bandwidth. Observations were made with a frequency resolution of 1.4\,kHz (translating to $0.05 - 0.01\,\mathrm{km/s}$ in velocity) and RMS noise between 2 and 20\,mK on the antenna temperature, varying with frequency due to integration times and atmosphere opacity corrections. Data reduction included calibrating to the internal noise diodes, removing interferences, subtracting baselines, and correcting for atmospheric attenuation and telescope efficiencies. The observational and general data reduction strategies are detailed in \citet{mcguire_early_2020} and \citet{sita_discovery_2022}, and an in depth description of recent advances in the data reduction will be found in Xue et al. (submitted). No transitions could be detected in the observational data for any of the molecules studied in this work, even using state-of-the-art spectral stacking and matched filtering techniques~\cite{loomis_investigation_2021}.

Based on the predicted intensities of rotational transitions and the noise levels in the observations, upper limits to the column densities of the five species toward TMC-1 were determined by calculating the minimum column density necessary for the detection of a single $3\,\sigma$ peak. For cyanocyclohexane, the axial and equatorial conformers were found to have upper limits of $2.00 \cdot 10^{12}\,\text{cm}^{-2}$ and $1.01 \cdot 10^{12}\,\text{cm}^{-2}$, respectively. An upper limit of $9.20 \cdot 10^{11}\,\text{cm}^{-2}$ was calculated for 1-cyanocyclohexene, and the upper limits of the axial and equatorial conformers of 4-cyanocyclohexene were determined to be $2.71 \cdot 10^{12}\,\text{cm}^{-2}$ and $1.45 \cdot 10^{12}\,\text{cm}^{-2}$, respectively.  

Many of these species possess favorable transitions between 35--50\,GHz, in the frequency range covered by the ultra-sensitive QUIJOTE (Q-band Ultrasensivite Inspection Journey to the Obscure TMC-1 Environment) survey of TMC-1 using the Yebes 40\,m telescope~\cite{cernicharo_discovery_2021-1}. The raw sensitivity of the QUIJOTE data and access to higher frequencies has enabled the detection of a number of single-ringed species in this source that were not optimally suited for discovery in the GOTHAM data~\cite{cernicharo_discovery_2021}. The spectroscopic fits and catalogs provided through this work should enable a robust search for these molecules in the QUIJOTE data.

\subsection{Astrochemical Implications}

The formation pathways of PAHs and their building blocks, such as benzene, in the interstellar medium are an active area of investigation in astrochemistry~\cite{jones_formation_2011}. However, many key, pure hydrocarbon species remain either invisible or difficult to detect with radio telescopes due to their low or nonexistent permanent electric dipole moments~\cite{mccarthy_laboratory_2006}. While cyclohexene does have a small dipole moment, it is effective to study nitrile-substituted molecules as analogs for their pure hydrocarbon counterparts. The addition of a -\ce{CN} group dramatically enhances the dipole moment of a molecule; the intensity of the observed rotational transitions goes as the square of the dipole, and thus very large gains in detectability can be made. Further, \ce{CN}-addition \ce{H}-elimination processes can be extremely efficient within TMC-1, and thus we studied nitrile functionalized versions of cyclohexenes as analogs to their pure hydrocarbon counterparts~\cite{cooke_benzonitrile_2020, wenzel_detection_2024}. 

Cyclohexene is an important intermediate compound along the pathways transforming cyclohexane into benzene~\cite{wang_understanding_2019}. Understanding these various formation steps may be key to solving the mystery of bottom-up PAH formation at low temperatures. The formation pathways of both benzene and cyclohexane have been extensively studied in combustion processes for several decades~\cite{bonner_cool_1965,mcenally_experimental_2004,lu_kinetic_2021}, but are more opaque in extremely low temperature environments more applicable to astrochemistry. Notably, cyclohexene is the most stable of the unsaturated six-membered rings~\cite{tsang_thermal_1978}. Along with other cyclic alkanes and alkenes, cyclohexene has been suggested as a possible culprit in the 3.4\,\textmu m spectral feature of infrared (IR) spectra of interstellar clouds and the diffuse ISM~\cite{duley_laboratory_2005}. 

Much of the relevant interstellar work investigating formation of cyclohexane or benzene has been done on icy surface analogs. \citet{pilling_formation_2012} studied the formation of unsaturated bonds in interstellar ice analogs by cosmic rays in cyclohexanes, finding the bombarded cyclohexane produced several aliphatic and aromatic unsaturated compounds including hexene, benzene, and importantly cyclohexene. They conclude that cosmic ray bombardment is a possible and maybe even likely pathway in the production of unsaturated hydrocarbons in interstellar ices. Meanwhile, surface hydrogenation of benzene is an area of high interest as a method of reducing aromatic bonds. Generally speaking, hydrogenation processes have a high activation energy barrier and in terrestrial environments require a catalyst~\cite{simons_formation_2020}. However, \citet{na_birch_2009} described the observation of benzene and arene hydrogenation in low-temperature plasma of helium at atmospheric pressure. They concluded that the detection of reduced compounds such as cyclohexadiene in their instrument relied upon surface adsorption in their experimental set-up; their helium plasma was hypothesized to negatively charge the surfaces, thus capturing low binding energy electrons onto the surfaces which could diffuse easily. These are especially relevant to interstellar studies, as grain surfaces are thought to be negatively charged and act as a reservoir of electrons~\cite{cui_exploring_2024}, such as in star-forming regions examined by the Spitzer telescope \cite{oberg_spitzer_2011}. 

Other studies of cyclohexene have focused specifically on their unique chemistry. Cyclohexene falls into the alkene class of compounds, which are reactive organic compounds. The hybridization of cyclohexene arising from the double bond makes it an attractive compound to pursue in the quest to understand cyclic molecule formation. \citet{chen_vacuum_2016} sought to study it as a surrogate to understand the photoionization and dissociation behavior of other alkenes. Their vacuum ultraviolet (VUV) photoionization mass spectrometry studies of cyclohexene found that it dissociates into several fragment species and ions. These include short chain species ($N_\mathrm{C}=4-6$), ring ions such as benzenium (\ce{C6H7+}), an important precursor to benzene formation~\cite{douberly_infrared_2008}, and species that have previously been detected in TMC-1, such as cyclopentadiene~\cite{cernicharo_pure_2021}. Of special interest is \ce{C6H7+}, which shows that further desaturation can occur from photoprocessing of cyclohexene.

Recent calculations by \citet{robertson_identifying_2021} aiming to identify barrierless mechanisms for benzene formation in the ISM from small hydrocarbon reactants with four or fewer carbon atoms suggested cyclohexenyl-like transition states and cyclohexadienes as probable intermediates within the barrierless formation pathways of benzene. Many of their results suggested that benzene could arise from small species forming a long chain, followed by an intramolecular cyclization step, rather than a Diels-Alder reaction-like concerted ring-closing reaction. Interestingly, while ring closure via a 5-hexenyl radical is a well-known and common method of forming five-membered rings in chemistry (often proving instrumental in the formation of species such as cyclopentadiene)~\cite{newcomb_competition_1993}, ring cyclization of 6-heptanyl radicals to make six-membered rings occurs at much slower rates due to the stereochemistry dependency~\cite{walling_cyclization_1972}.

Much of the previous discussion has focused on studies in environments similar to those in the ISM. However, most research on cyclohexene formation in the gas phase has been conducted in the context of terrestrial chemistry, which may not be directly applicable to astrophysical environments. For example, \citet{shao_thermal_2021} studied the thermal decomposition of cyclohexane up to ${\sim}1300\,\mathrm{K}$ in a vacuum, finding that \ce{H2} elimination from cyclohexane into cyclohexene is a secondary route. As well, studies of the oxidation and auto-ignition of cyclohexane, cyclohexene, and cyclohexa-1,3-diene were studied in the $600-900\,\mathrm{K}$ range by  \citet{lemaire_production_2001}, showing that hydrogen abstraction by oxygen species is a primary pathway in the formation of cyclohexene from cyclohexane. While this is unlikely to occur in cold dark regions such as TMC-1, other pathways of hydrogen abstraction could occur, as discussed above. 
 
We reiterate that these last few mechanisms were proposed for terrestrial/combustion studies. While it is difficult to confidently assert that these reactions would or could occur in the ISM, it is worth noting that if an alternative pathway with a lower energy barrier, such as hydrogen abstraction via a different reaction sequence, can be identified, it may offer a promising avenue for further investigation. As it stands, hydrogen abstraction as a standalone process presents a high energy barrier that would be challenging to overcome~\cite{gong_theoretical_2012}.

\section{Conclusions}

The development of a compact CP-FTMW spectrometer operating in the range of $6.5-18\,\mathrm{GHz}$ has enabled investigations of cyanocyclohexane, 1-cyanocyclohexene, and 4-cyanocyclohexene, the latter of which had not been previously studied using rotational spectroscopy. The broad bandwidth, high sensitivity, and ability to resolve hyperfine splitting from nuclear quadrupole coupling have provided rotational spectra from which detailed structural insights and the conformational landscape of these molecules could be derived, allowing us to determine the structure of 4-cyanocyclohexene. These laboratory measurements supply accurate spectroscopic parameters for astronomical searches of these species in the ISM.

While no detections were made searching toward TMC-1 CP, our results provide valuable upper limits on the column densities and suggest that the formation and survival of cyanocyclohexane derivatives under interstellar conditions may be limited. Future studies could focus on complementary environments, such as warm or photochemically active regions, to explore the potential presence of these molecules in other astrochemical contexts.

\subsubsection*{Biographies}

Gabi Wenzel is a Research Scientist at the Department of Chemistry at the Massachusetts Institute of Technology. She earned her Bachelor of Science and Master of Science degrees in Physics from the University of M\"{u}nster, Germany, in 2014 and 2017, respectively, and her PhD in Laboratory Astrophysics at the University of Toulouse, France, in 2020, before gaining first postdoctoral research experience at the University of Aarhus, Denmark, until 2022. Her research focuses on the detection of cosmic PAHs in space using rotational spectroscopy and radio astronomy. She further works on understanding their photophysics, relaxation channels, and possible formation pathways in the interstellar medium.

Martin S. Holdren is a postdoctoral associate in the Chemistry Department at the Massachusetts Institute of Technology. He received his B.S. in Chemistry and Applied Mathematics from Mansfield University of Pennsylvania in 2016 and his Ph.D. in Analytical Chemistry from the University of Virginia in 2022. Martin’s research background is in the development of instrumentation and strategies using rotational spectroscopy to tackle analytical chemistry problems in impurity analysis, chiral analysis, and isotopic isomer analysis relevant to pharmaceutical manufacturing, environmental sciences, and astrochemistry. 

D. Archie Stewart earned his B.S. in Chemistry and Applied Mathematics from Emory University in 2022 and is currently pursuing a PhD in chemistry at the Massachusetts Institute of Technology. His current research investigates the chemistry of substituted benzene species in the ISM with laboratory rotational spectroscopy and radioastronomical observations. Further research interests focus on using machine learning models to analyze rotational spectra.

Hannah Toru Shay is currently pursuing a PhD in chemistry at the Massachusetts Institute of Technology. She earned her master's in Chemistry from the University of Illinois at Urbana-Champaign in 2021. Her current work explores the intersections of chemistry, machine learning and data science methods. Other research interests include spectroscopy, computational chemistry, astrochemistry and environmental chemistry.

Alex N. Byrne obtained his B.S. in Chemistry from Temple University in 2021 and is now a PhD candidate in the Chemistry department at the Massachusetts Institute of Technology. His research involves the use of computational chemistry techniques such as kinetic simulations to study the formation of complex organic molecules in the interstellar medium. In particular, he is interested in understanding the formation of PAHs in cold molecular clouds and the overlap with combustion and atmospheric chemistry.

Ci Xue is is a postdoctoral associate in the Chemistry Department at the Massachusetts Institute of Technology. She received her B.S in Biotechnology from Xiamen University, China, in 2015, and her Ph.D. in Physical Chemistry from the University of Virginia in 2021. She has worked as a research intern at Academia Sinica Institute of Astronomy and Astrophysics, Taiwan, and Max Planck Institute for Astronomy, Germany. Her research focuses on using radio astronomy observations in conjunction with spectral modeling to study interstellar chemistry.

Brett A. McGuire is an Assistant Professor of Chemistry at the Massachusetts Institute of Technology. He received his B.S. in Chemistry from the University of Illinois in 2009 and his Ph.D. in Physical Chemistry from Caltech in 2015. He was a Jansky and then Hubble Postdoctoral Fellow at the National Radio Astronomy Observatory from 2014--2020.  Research in McGuire's group uses the tools of rotational spectroscopy, observational astronomy, computational chemistry, and machine learning to understand the evolution of molecular complexity along the path of star- and planet-formation.

\begin{acknowledgement}

We are grateful to Francis Lovas (NIST) and Robert W. Field (MIT) for donating equipment to this work and our lab in general. We thank Brooks H. Pate (UVA) for discussions and his expertise on the new circuit and Nigel Atkins (CfA) for his input on electronics. We also thank Kyle Crabtree (UC Davis) for making \textsc{blackchirp} an open-source software. G.W., D.A.S., and B.A.M. acknowledge support from an Arnold and Mabel Beckman Foundation Beckman Young Investigator Award. M.S.H. and B.A.M. acknowledge support from the Schmidt Family Futures Foundation. A.N.B. acknowledges support from NSF Graduate Research Fellowship grant 2141064. C. X. and B.A.M. acknowledge support of National Science Foundation grant AST-2205126.\\

\end{acknowledgement}


\providecommand{\latin}[1]{#1}
\makeatletter
\providecommand{\doi}
  {\begingroup\let\do\@makeother\dospecials
  \catcode`\{=1 \catcode`\}=2 \doi@aux}
\providecommand{\doi@aux}[1]{\endgroup\texttt{#1}}
\makeatother
\providecommand*\mcitethebibliography{\thebibliography}
\csname @ifundefined\endcsname{endmcitethebibliography}
  {\let\endmcitethebibliography\endthebibliography}{}

\clearpage

\setcounter{page}{1}
\onecolumn

\vspace{1.5cm}

\begin{suppinfo}

\begin{center}

\vspace{1.5cm}

\textbf{\huge Laboratory Rotational Spectra of Cyanocyclohexane and
its Siblings (1- and 4-Cyanocyclohexene) Using a Compact
CP-FTMW Spectrometer for Interstellar Detection}

\vspace{1.5cm}

\textbf{Gabi Wenzel,$^{1,\star}$ Martin S. Holdren,$^{1,\star}$ D. Archie Stewart,$^{1}$ Hannah Toru Shay,$^{1}$\\ Alex N. Byrne,$^{1}$ Ci Xue,$^{1}$ and Brett A. McGuire$^{1,2,\star}$}\\
\vspace{0.5cm}

\textit{$^{1}$ Department of Chemistry, Massachusetts Institute of Technology, Cambridge, MA 02139, USA}\\
\textit{$^{2}$ National Radio Astronomy Observatory, Charlottesville, VA 22903, USA}

\vspace{0.5cm}

{$^\star$E-mail: gwenzel@mit.edu; holdrenm@mit.edu; brettmc@mit.edu}\\
\vspace{1.5cm}

\end{center}

{\noindent\Large\textbf{Contents}}

\renewcommand{\thefigure}{S\arabic{figure}}
\renewcommand{\thetable}{S\arabic{table}}
\renewcommand{\theequation}{S\arabic{equation}}
\renewcommand{\thesection}{S\arabic{section}}
\renewcommand{\thepage}{S\arabic{page}}
\setcounter{figure}{0}
\setcounter{table}{0}
\setcounter{equation}{0}
\setcounter{section}{0}
\setcounter{page}{1}

\textbf{
\begin{itemize}
    \item[S1] Parts of the CP-FTMW Spectrometer
    \item[S2] Heated Nozzle
    \item[S3] Optimized Geometries
    \item[S4] Expanded Quantum Calculations for 4-Cyanocyclohexene
    \item[S5] Rotational Spectrum of OCS
    \item[S6] Full Spectrum of 1-Cyanocyclohexene
    \item[S7] Full Spectrum of 4-Cyanocyclohexene
    \item[S8] Experimentally Scaled Rotational Constants of 1- and 4-Cyanocyclohexene Isotopic Isomers
    \item[S9] Kraitchman Coordinates: Experimental vs. Calculated
\end{itemize}}

\clearpage

\section{Parts of the CP-FTMW Spectrometer}

\begin{figure}[htb!]
    \centering
    \includegraphics[width=0.6\textwidth]{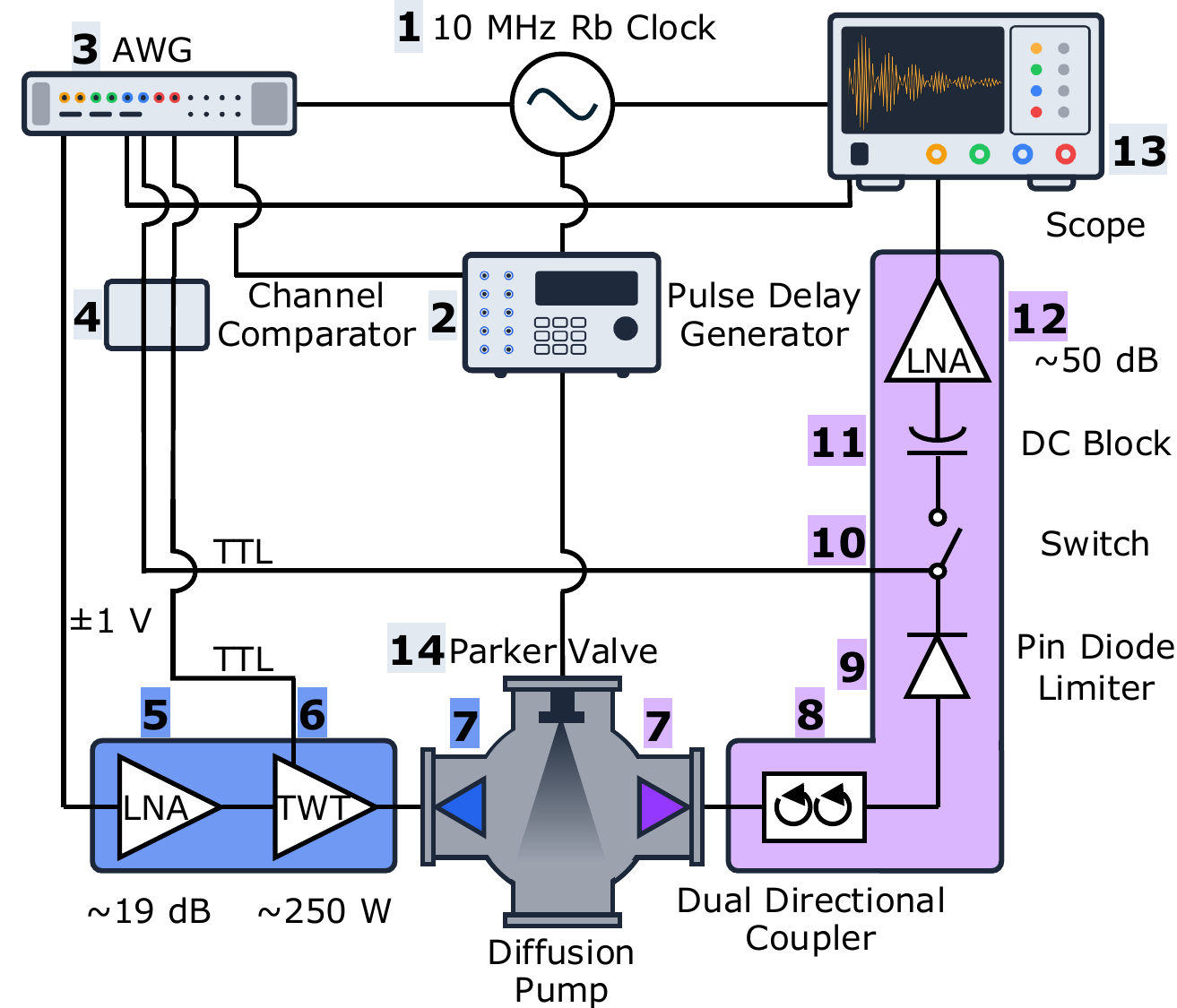}
    \caption{Schematic block diagram of the CP-FTMW spectrometer operating from $6.5$ to $18.0\,\mathrm{GHz}$. Blue color marks the excitation, purple color the receiver side of the spectrometer circuit. All parts are listed in Table~\ref{tab:parts}.}
    \label{fig:SI_tweety_circuit}
\end{figure}

\begin{table*}[htb!]
    \centering
    \begin{tabular}{clll}
    \toprule
         \textbf{Number} & \textbf{Manufacturer} & \textbf{Part Number} & \textbf{Notes} \\
         \midrule
        1 & Stanford Research & FS725 & 10\,MHz Rubidium Frequency Standard,\\
         & Systems &  & 8 outputs \\
         -- & Stanford Research & DS345 & Synthesized Function Generator, 30\,MHz, \\
          & Systems &  & 1\,$\upmu$Hz frequency resolution \\
        2 & Quantum Composer & QC9528 & Delay Pulse Generator, \\
         &  &  & jitter $<$50\,ps, 8 outputs \\
        3 & Keysight & M8195A & Arbitrary Waveform Generator, 65\,GS/s,\\
         &  &  & 25\,GHz, 4 channels\\
        4 & Pulse Research Lab & PRL-350TTL & Channel Comparator, 2 channels \\
        5 & RF Lambda & R06G18GSA & Low Noise Amplifier, 6--18\,GHz,\\
         &  &  & 17--20\,dB gain, 1.8\,dB noise figure\\
        6 & Applied Systems & 367X/Ku TWT & Traveling-Wave Tube Amplifier, \\
          & Engineering & & 6.5--18\,GHz, 250\,W\\
        7 & Microtech & HWRD650 & Standard Gain Horns, 6.5--18\,GHz \\
        8 & RF Lambda & RFDDC2G18G30 & Dual Directional Coupler, 2--18\,GHz\\
        9 & Pasternack & PE80L3000 & Pin Diode Limiter, 2--18\,GHz,\\
         &  &  &  200\,W Peak Power\\
        10 & RF Lambda & RFSPSTA0118G & SPST Switch, 1--18\,GHz\\
        11 & RF Lambda & RFDCBLK26SMA & DC Block, DC--26.5\,GHz\\
        12 & RF Lambda & RLNA06G18G45 & Ultra Low Noise Amplifier with Heatsink,\\
         &  &  &  6--18\,GHz\\
        13 & Keysight & DSOV204A & Oscilloscope, 80\,GS/s, 20\,GHz, 4 channels \\
        -- & Parker & 060-0001-900 & Iota One Valve Driver \\
        14 & Parker &  009-1643-900 & Series 9 Solenoid Valve \\
        \bottomrule
    \end{tabular}
    \caption{Component list for the CP-FTMW spectrometer depicted in Figs.~\ref{fig:tweety_circuit} and \ref{fig:SI_tweety_circuit}.}
    \label{tab:parts}
\end{table*}

\clearpage
\section{Heated Nozzle}
\label{sec:nozzle}
The measurement of a molecule’s rotational spectrum requires having a sufficient amount of the analyte in the gas phase that can be mixed with a carrier gas in the pulsed supersonic jet expansion technique commonly used to rotationally cool the analyte. While measurement conditions can be different for each experiment, typically a mixture of 0.1-1\,\% analyte in carrier gas (He, Ar, Ne) is used, with absolute pressure of the carrier gas around 1-4\,atm~\cite{neill_analysis_2023}. For very small molecules, including most of those that have been detected in the ISM already, this requirement tends to be trivial as these molecules are typically either a gas or a liquid with high vapor pressure at standard room temperature and pressure. However, larger and more complex molecules that are now becoming targets for searches in space often have much lower vapor pressures and may even be solids at standard room temperature and pressure. In order to measure the rotational spectra of these species, several research groups have designed elegant ways to increase the vapor pressure of an analyte in a controlled manner without appreciable molecular decomposition. Akin to instruments in the literature (Grubbs~\cite{sedo_rotational_2019}, Suenram~\cite{suenram_reinvestigation_2001},
Schnell~\cite{schmitz_multi-resonance_2012},
Jaeger~\cite{hazrah_structure_2022},
Cernicharo~\cite{cabezas_laboratory_2023},
McCarthy~\cite{thorwirth_rotational_2005},
Pate~\cite{neill_online_2019}, and more), in this work, a sample holding reservoir capable of being heated was custom-built to attach to a commercially available Parker Series 9 Solenoid Valve (PN: 009-1643-900) controlled with a Parker Iota One Valve Driver (PN: 060-0001-900). 

\begin{figure}[htb!]
    \centering
    \includegraphics[width=\textwidth]{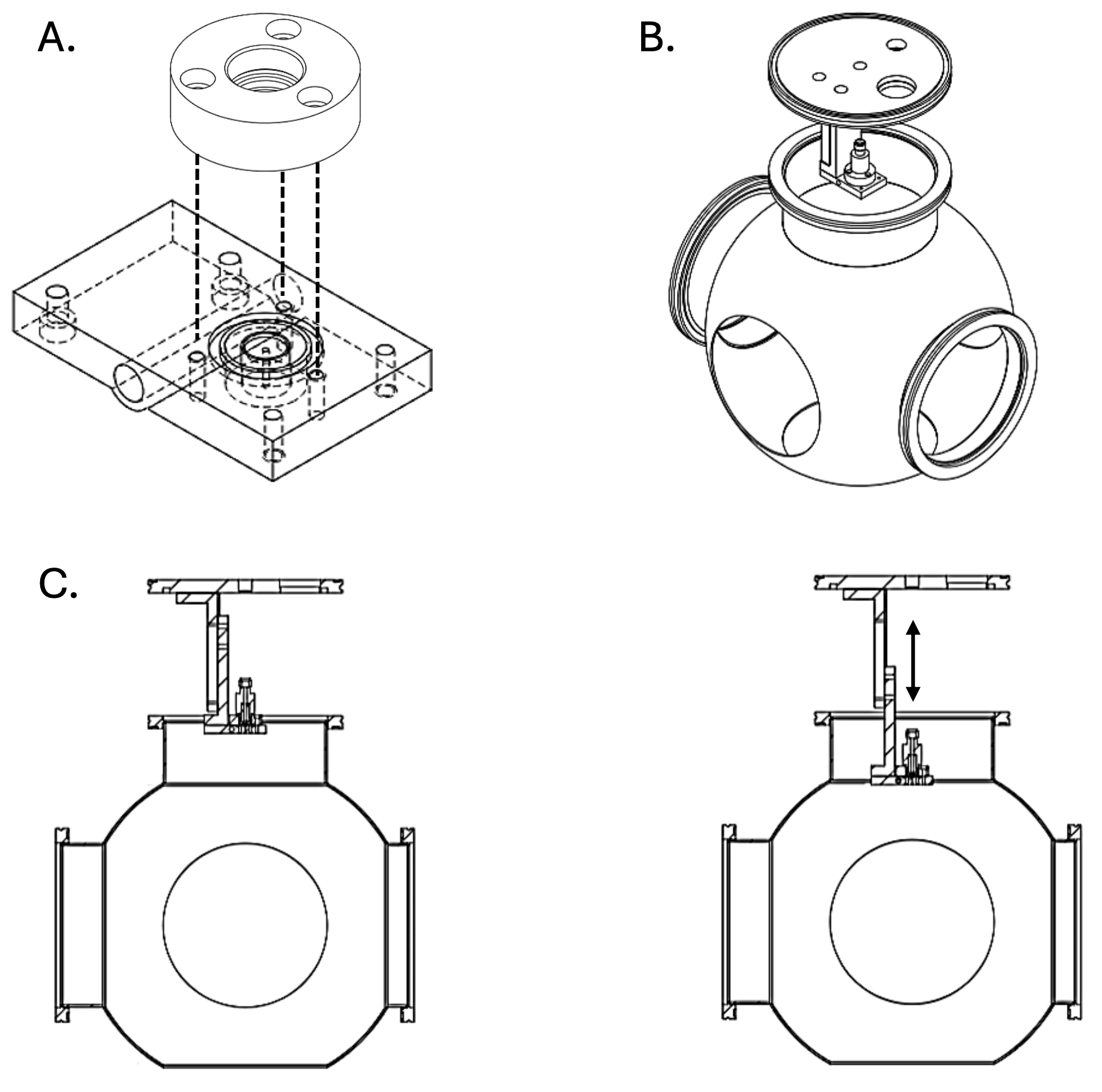}
    \caption{(A) Stainless steel baseplate attachment for the heated nozzle design containing the well reservoir for liquid and solid samples. A circular stainless steel piece is used to mate the commercial Parker solenoid valve to the baseplate. (B) Mounting of the baseplate within the vacuum chamber. The exterior steel jacket around the nozzle assembly and the gas tubing running through the center of the solenoid and baseplate are excluded here for better viewing. (C) Display of the vertical movement with in the vacuum chamber allowed by the design. The lower L-shaped bracket connecting the baseplate to the chamber is made of macor ceramic.}
    \label{fig:Heated-sample-holder}
\end{figure}

Typically, the details of the sample reservoirs used in the literature are not discussed, so a detailed description of this work’s reservoir is provided, and the CAD drawings are given. The stainless-steel body of this reservoir contains a hollowed-out well in a ring around the center piece where the standard nozzle assembly (armature, springs, and poppet) rests. This raised center piece also contains the 1\,mm orifice through which gas passes into the chamber. Liquid or solid sample may be deposited in the hollowed out well, with a volume ${\sim}0.5\,\mathrm{mL}$ (although this volume can be increased in the future). A cartridge heater (Omega Engineering Inc., PN: HDC00015) and thermocouple (Omega Engineering Inc., PN: WTK-8-12-TT) are slotted into the body and these are fed outside the vacuum chamber to a heater controller (Omega Engineering Inc., PN: CN733). The heating of this nozzle is currently limited to ${\sim}200^\circ\,\mathrm{C}$ with the main restriction being the temperature rating of the solenoid valve. Attached to the body is a stainless-steel ring used (with o-ring seal) to mate the sample holder to the solenoid valve. The stainless-steel body is then fixed in place to the chamber by a ceramic bracket with low thermal conductivity, which is able to slide the whole assembly up and down. This ceramic bracket allows the reservoir to be more thermally isolated during heating so that the main route for cooling is through the valve and metal gas line tubing, which in turn keeps the analyte from condensing on any cold surfaces or clogging the pulsed valve. Fixing the nozzle in place is necessary so that the tensioning between the reservoir and solenoid valve can be adjusted by rotating the sample introduction gas line outside the chamber to tighten or loosen this connection. As the equipment heats, the metal pieces of the pulsed valve assembly expand, and adjusting this connection’s tensioning can significantly change the free jet expansion to increase detected signals. 

Moving forward, the development of a more universal nozzle is of interest as there are several experiments that would require the combination of controlled heating followed by a discharge or perhaps a separate laser ablation or pyrolysis attachment for samples that have far lower vapor pressures; this universal nozzle would allow for quick exchange of samples such that entire assemblies need not be created for each new sample. 

\clearpage

\section{Optimized Geometries}

\begin{figure}[hbt!]
    \centering
    \includegraphics[width=0.5\textwidth]{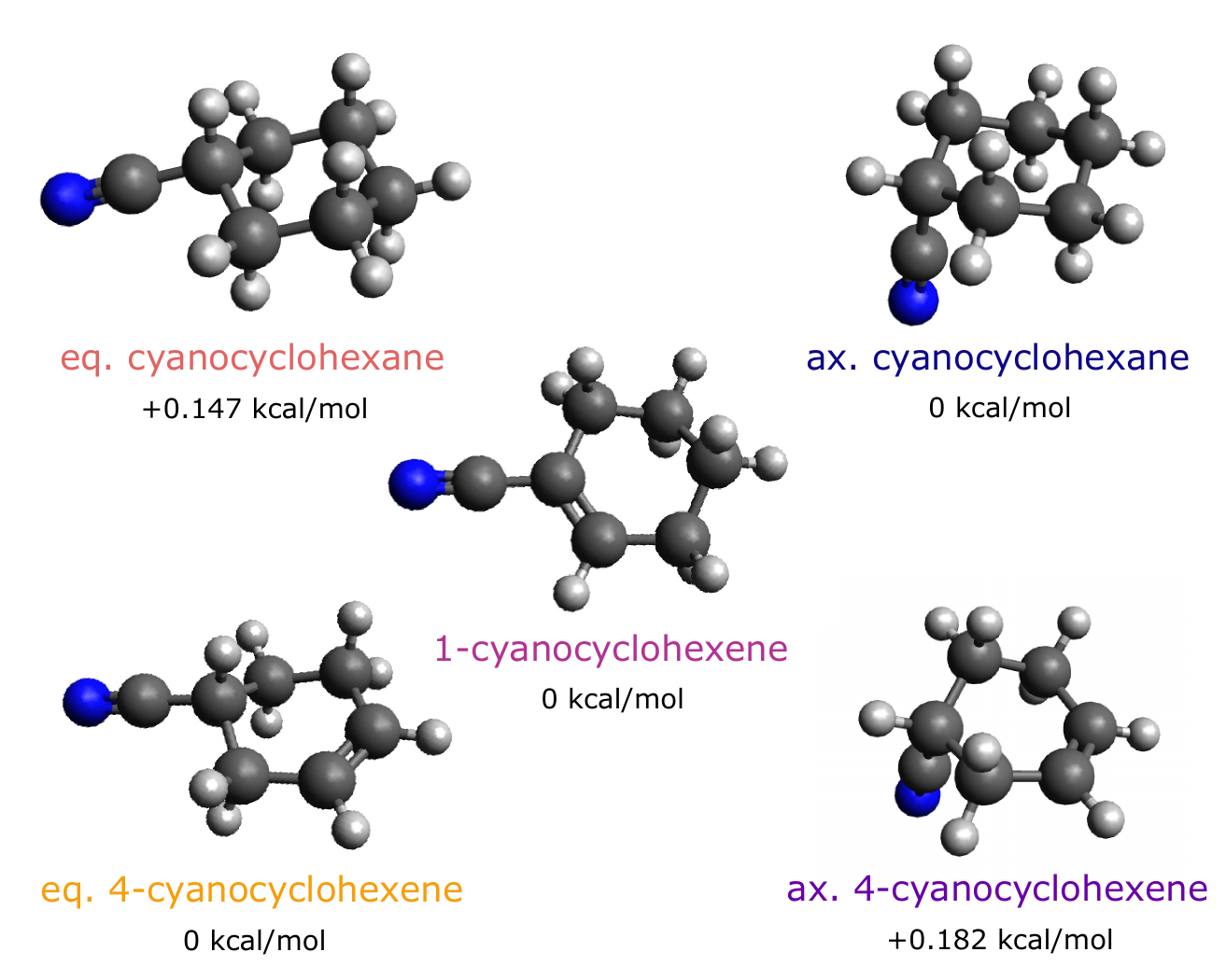}
    \caption{Optimized geometries at the B3LYP-D3(BJ)/6-311++G(d,p) level of theory of the present molecules. Top row cyanocyclohexane in chair-equatorial and chair-axial conformation, middle 1-cyanocyclohexene as its one and only chair-equatorial conformer, and bottom row 4-cyanocyclohexene in its chair-equatorial and chair-axial conformation. Relative energies between the axial and equatorial conformers is noted below each pair in 1- and 4-cyanocyclohexene.}
    \label{fig:mol_structures}
\end{figure}

\section{Expanded Quantum Calculations for 4-Cyanocyclohexene}

\begin{table*}[htb!]
    \centering
    \caption{Experimentally fit spectroscopic constants for axial 4-cyanocyclohexene, {ax. \ce{C6H9}-\ce{4-CN}}, and equatorial 4-cyanocyclohexene, {eq. \ce{C6H9}-\ce{4-CN}}, compared to calculations using alternative levels of theory. All spectroscopic parameters are in MHz.}
    \label{tab:4cn_ax_eq_extra}
    \begin{threeparttable}
    \begin{tabular}{llS[table-format=3.10]S[table-format=3.5]S[table-format=3.5]S[table-format=3.5]}
        \multicolumn{6}{c}{\textbf{axial \ce{C6H11-CN}}}\\
    \toprule
        && & {B2PLYPD3} & {B2PLYPD3}& {B3LYP-D3(BJ)} \\
        
        {Parameter} && {Experimental} & {6-311++G(d,p)} & {def2-TZVP} & {def2-TZVP} \\
  \cline{1-1}
    \cline{3-6}
        & & &  &  & \\
        $A$ && 2982.81120(50) & 2997.6543 & 3011.2351 & 3028.1042 \\
        $B$ && 1899.16757(39) & 1884.8859 & 1896.4566 & 1885.6910  \\
        $C$ && 1696.22518(42) & 1679.4249 & 1690.0646 & 1674.6369  \\
        $\chi_{aa}$ && -0.9657(56) & -1.0087 & -1.0459 & -1.1424  \\
        $\chi_{bb}$ && 1.9983(51) & 1.9866 & 2.1002 & 2.2590 \\
        \bottomrule
    & \\
        \multicolumn{6}{c}{\textbf{equatorial \ce{C6H11-CN}}}\\
    \toprule
        && & {B2PLYPD3} & {B2PLYPD3}& {B3LYP-D3(BJ)} \\
        
        {Parameter} && {Experimental} & {6-311++G(d,p)} & {def2-TZVP} & {def2-TZVP} \\
  \cline{1-1}
    \cline{3-6}
        & & &  &  & \\
        $A$ && 4561.31154(92) & 4564.9032 & 4589.0991 & 4582.7884 \\
        $B$ && 1460.14323(49) & 1456.5333 & 1464.5925 & 1464.8020 \\
        $C$ && 1160.76780(41) & 1157.9954 & 1164.3520 & 1163.6081 \\
        $\chi_{aa}$ && -4.0066(41) & -3.9668 & -4.1924 & -4.4825 \\
        $\chi_{bb}$ && 2.1228(54) & 2.0904 & 2.2091 & 2.3696 \\

        \bottomrule
    \end{tabular}
    \end{threeparttable}
\end{table*}

\clearpage

\section{Rotational Spectrum of OCS}

\begin{figure*}[htb!]
    \centering
    \includegraphics[width=\linewidth]{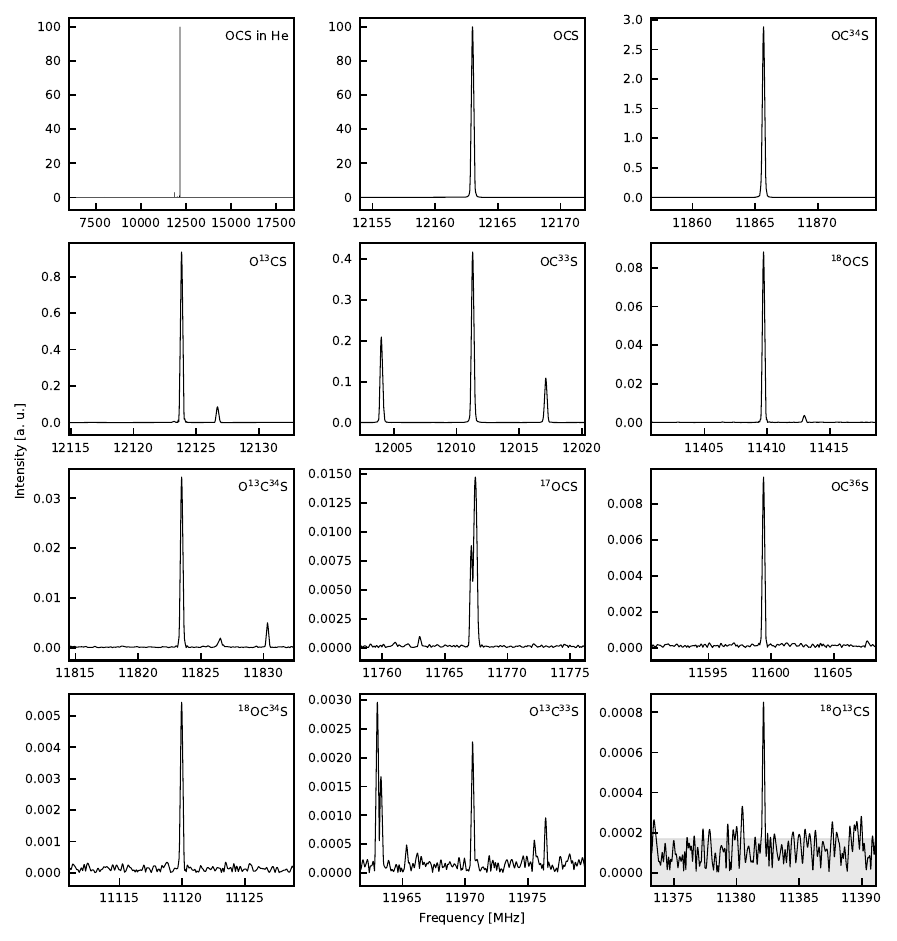}
    \caption{Experimental spectrum of 0.5\,\% \ce{OCS} diluted in He at 1\,M shots. The spectrum has been normalized to 100 on the $J = 1 - 0$ transition of \ce{OCS} (\ce{^{16}O^{12}C^{32}S}) at 12162.977\,MHz for better visual interpretation and comparison. Isotopic isomers are observable down to a 20\,ppm level in natural abundance (see Table~\ref{tab:ocs}) and depicted in the panels. The noise level at approximately 0.05\,$\upmu$V is shown in grey in the bottom right panel.}
    \label{fig:ocs_spectra}
\end{figure*}

\begin{table*}[htb!]
    \centering
    \caption{List of the $J = 1-0$ rotational transitions of isotopic isomers of \ce{OCS} measured in natural abundance (see Fig.~\ref{fig:ocs_spectra}) and comparison of frequency accuracy to measurements by the NIST FTMW cavity spectrometer.\cite{lovas_pulsed_1987}}
        \label{tab:ocs}
    \begin{threeparttable}
    \begin{tabular}{lccS[table-format=3.5]S[table-format=3.5]S[table-format=3.5]S[table-format=3.5]S[table-format=3.5]S[table-format=3.5]}
    
    \toprule      
        {Isotopologue} & {$F$} & & {Measured} & {NIST\cite{lovas_pulsed_1987}} & & {Measured} & {Measured} & {NIST\cite{lovas_pulsed_1987}}\\
        
         &  &  & {Freq.} & {Freq.} & & {Intensity} & {Nat. Abund.} & {Nat. Abund.}\\

         &  &  & {(MHz)} & {(MHz)} & & {(µV)}\tnote{a,b,c}  & {(\%)} & {(\%)}\\
         
    \midrule
        OCS &  & & 12162.9774 & 12162.9790 &  & 29323.9884 & 95.52 & 93.74 \\
        OC${}^{34}$S & &  & 11865.6685 & 11865.6628 &  & 835.0869 & 2.720 & 4.158  \\
        O${}^{13}$CS & &  & 12123.8290 & 12123.8420 &  & 274.3324 & 0.894 & 1.053 \\
        OC${}^{33}$S & $\text{3/2 } - \text{3/2}$&  & 12003.9858 & 12004.0029 & & 61.4110 & 0.200 & 0.24  \\
         & $\text{5/2 } - \text{3/2}$&  & 12011.2768 & 12011.2849 & & 122.4230 & 0.399 & 0.37 \\
         & $\text{1/2 } - \text{3/2}$&  & 12017.1125 & 12017.1050 & & 31.8728 & 0.104 & 0.13 \\
        ${}^{18}$OCS &  &&  11409.7199 & 11409.7097 & &  25.8398 & 0.0842 & 0.188 \\
        O${}^{13}$C${}^{34}$S &  & & 11823.4634 & 11823.4625 & & 10.0468 & 0.0327 & 0.0467 \\
        ${}^{17}$OCS & $\text{5/2 }-\text{5/2}$ & & 11767.1018 & 11767.1315 & & 2.5826 & 0.00841 & 0.012  \\
         & $\text{7/2 } - \text{5/2}$ & & 11767.4209 & 11767.4015 & & 2.1584 & 0.00703 & 0.015 \\
         & $\text{3/2 } - \text{5/2}$ & & 11767.4209 & 11767.5330 & & 2.1584 & 0.00703 & 0.008 \\
        OC${}^{36}$S & &  & 11599.3983 & 11599.3816 & & 2.7833 & 0.00907 & 0.0167 \\
        ${}^{18}$OC${}^{34}$S & &  & 11119.9534 & 11119.9346 & & 1.5942 & 0.00519 & 0.00834 \\
        O${}^{13}$C${}^{33}$S & $\text{3/2 } - \text{3/2}$ &&  11963.2692 & 11963.3002 & & 0.4759 & 0.00155 & 0.00274  \\
         & $\text{5/2 } - \text{3/2}$ &&  11970.5796 & 11970.5845 & & 0.6668 & 0.00217 & 0.00416 \\
         & $\text{1/2 } - \text{3/2}$ &&  11976.4093 & 11976.4055 & & 0.2804 & 0.000913 & 0.00141 \\
        ${}^{18}$O${}^{13}$CS & &  & 11382.1590 & 11382.1280 & & 0.2493 & 0.000812 & 0.00211 \\
        \bottomrule
    \end{tabular}
        \begin{tablenotes}
        \item[a] Corrections in measured signal intensities for small deviations of the electric dipole moment for each isotopomer are not included.
        \item[b] Corrections in measured signal intensities for small deviations of the power spectrum as a function of frequency are not included.
        \item[c] Noise level at 1\,M shots is 0.05 µV.
    \end{tablenotes}
    \end{threeparttable}
\end{table*}

\clearpage

\section{Full Spectrum of 1-Cyanocyclohexene}

\begin{figure*}[htb!]
    \centering
    \includegraphics[width=\linewidth]{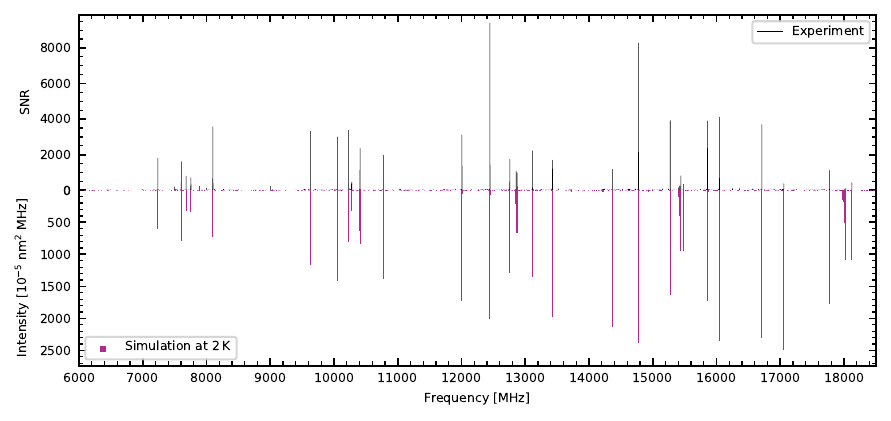}
    \caption{Experimental spectrum of 1-cyanocyclohexene seeded in Ar at 1.5\,M shots in comparison to the fit on the inverted $y$-axis using the spectroscopic constants presented in Table~\ref{tab:1cchexene}.}
    \label{fig:1cchexene_full}
\end{figure*}

\section{Full Spectrum of 4-Cyanocyclohexene}

\begin{figure*}[htb!]
    \centering
    \includegraphics[width=\textwidth]{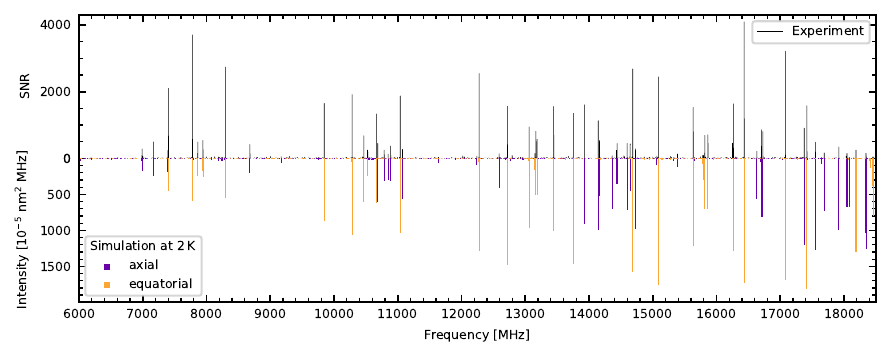}
    \caption{Experimental spectrum of 4-cyanocyclohexene seeded in Ar at 1.5\,M shots in comparison to the fits of the axial (violet) and equatorial (yellow) conformers on the inverted $y$-axis using the spectroscopic constants presented in Table~\ref{tab:4cchexene_ax_eq}.}
    \label{fig:4cchexene_full}
\end{figure*}

\clearpage

{
\section{Experimentally Scaled Rotational Constants of 1- and 4-Cyanocyclohexene Isotopic Isomers} }

\begin{table*}[htb!]
    \centering
    \caption{{Calculated spectroscopic constants for the isotopic isomers of 1-cyanocyclohexene, {\ce{C6H9}-\ce{1-CN}} at the B3LYP-D3(BJ)/6-311G++(d,p) level of theory. These constants are then scaled based on the error between calculation and experimental fit for each rotational constant of the parent isotopologue species. All spectroscopic parameters are in MHz.}}
    \label{tab:1cchexene_isos_calc}
    \begin{threeparttable}
    \begin{tabular}{lcccccccc}
      &  \multicolumn{8}{c}{\textbf{\ce{C6H9}-\ce{1-CN}}}\\
    \toprule
            && \multicolumn{3}{c}{Calculated} && \multicolumn{3}{c}{Scaled} \\
        {Species} && $A$ & $B$ & $C$ && $A$ & $B$ & $C$ \\
    \cline{1-1}
    \cline{3-5}
    \cline{7-9}
        && & & && & & \\
        {${}^{13}C_1$} && 4573.962 & 1422.792 & 1134.898 && 4565.712 & 1421.932 & 1135.120 \\
        {${}^{13}C_2$} && 4514.220 & 1424.273 & 1132.167 && 4506.078 & 1423.412 & 1132.469  \\
        {${}^{13}C_3$} && 4505.729 & 1414.755 & 1125.612 && 4497.602 & 1413.900 & 1125.911 \\
        {${}^{13}C_4$} && 4568.995 & 1404.549 & 1123.572 && 4560.754 & 1403.700 & 1123.871 \\
        {${}^{13}C_5$} && 4511.332 & 1414.461 & 1126.198 && 4503.195 & 1413.606 & 1126.497 \\
        {${}^{13}C_6$} && 4505.148 & 1424.273 & 1131.586 && 4497.022 & 1423.412 & 1131.887  \\
        {${}^{13}C_7$} && 4573.953 & 1407.766 & 1125.319 && 4565.704 & 1406.916 & 1125.618 \\
        {${}^{15}$N} && 4573.992 & 1384.736 & 1110.563 && 4565.742 & 1383.899 & 1110.858 \\
        \bottomrule
    \end{tabular}
    \end{threeparttable}
\end{table*}

\begin{table*}[htb!]
    \centering
    \caption{Calculated spectroscopic constants for the isotopic isomers of axial 4-cyanocyclohexene, {ax. \ce{C6H9}-\ce{4-CN}}, and equatorial 4-cyanocyclohexene, {eq. \ce{C6H9}-\ce{4-CN}}, at the B3LYP-D3(BJ)/6-311G++(d,p) level of theory. These constants are then scaled based on the error between calculation and experimental fit for each rotational constant of the parent isotopologue species. All spectroscopic parameters are in MHz.}
    \label{tab:4cchexene_ax_eq_isos_calc}
    \begin{threeparttable}
    \begin{tabular}{lcccccccc}
      &  \multicolumn{8}{c}{\textbf{axial {\ce{C6H9}-\ce{4-CN}}}}\\
    \toprule
            && \multicolumn{3}{c}{Calculated} && \multicolumn{3}{c}{Scaled} \\
        {Species} && $A$ & $B$ & $C$ && $A$ & $B$ & $C$ \\
    \cline{1-1}
    \cline{3-5}
    \cline{7-9}
        && & & && & & \\
        {${}^{13}C_1$} && 3000.423 & 1855.394 & 1653.908 && 2969.996 & 1875.585 & 1681.196 \\
        {${}^{13}C_2$} && 2988.778 & 1864.660 & 1652.446 && 2958.469 & 1884.951 & 1679.710  \\
        {${}^{13}C_3$} && 2971.833 & 1875.929 & 1656.981 && 2941.696 & 1896.343 & 1684.320 \\
        {${}^{13}C_4$} && 2998.021 & 1870.620 & 1666.446 && 2967.619 & 1890.976 & 1693.942 \\
        {${}^{13}C_5$} && 2974.194 & 1870.926 & 1662.030 && 2944.033 & 1891.286 & 1689.452 \\
        {${}^{13}C_6$} && 2979.004 & 1871.313 & 1652.948 && 2948.794 & 1891.677 & 1680.221  \\
        {${}^{13}C_7$} && 3013.066 & 1860.428 & 1654.315 && 2982.511 & 1880.673 & 1681.611 \\
        {${}^{15}$N} && 2998.411 & 1832.988 & 1636.813 && 2968.005 & 1852.935 & 1663.820 \\
        \bottomrule
        & \\
            & \multicolumn{8}{c}{\textbf{equatorial {\ce{C6H9}-\ce{4-CN}}}}\\
    \toprule
            && \multicolumn{3}{c}{Calculated} && \multicolumn{3}{c}{Scaled} \\
        {Species} && $A$ & $B$ & $C$ && $A$ & $B$ & $C$\\
    \cline{1-1}
    \cline{3-5}
    \cline{7-9}
        && & & && & &  \\
        {${}^{13}C_1$} && 4561.811 & 1437.974 & 1145.582 && 4560.938 & 1439.259 & 1147.523 \\
        {${}^{13}C_2$} && 4501.315 & 1448.542 & 1148.403 && 4500.453 & 1449.837 & 1150.349  \\
        {${}^{13}C_3$} && 4492.658 & 1458.772 & 1154.287 && 4491.797 & 1460.077 & 1156.243 \\
        {${}^{13}C_4$} && 4556.738 & 1456.905 & 1157.934 && 4555.865 & 1458.207 & 1159.896  \\
        {${}^{13}C_5$} && 4498.462 & 1458.395 & 1154.853 && 4497.600 & 1459.699 & 1156.810 \\
        {${}^{13}C_6$} && 4497.921 & 1448.140 & 1147.948 && 4497.059 & 1449.435 & 1149.893  \\
        {${}^{13}C_7$} && 4561.622 & 1442.006 & 1148.192 && 4560.749 & 1443.295 & 1150.137 \\
        {${}^{15}N$} && 4561.782 & 1418.735 & 1133.379 && 4560.909 & 1420.004 & 1135.299 \\
        \bottomrule
    \end{tabular}
    \end{threeparttable}
\end{table*}

\clearpage

\section{Kraitchman Coordinates: Experimental vs. Calculated}

\begin{table*}[h!]
    \centering
    \caption{Experimentally determined Kraitchman coordinates of the heavy atoms of 1-cyanocyclohexene, {\ce{C6H9}-\ce{1-CN}}, compared to those calculated at the B3LYP-D3(BJ)/6-311G++(d,p) level of theory. Coordinates are given in Angstroms (\r{A}).}
    \label{tab:1-cchexene_kraitchman_coords}
    \begin{threeparttable}
    \begin{tabular}{lcS[table-format=3.6]S[table-format=3.5]S[table-format=3.5]cS[table-format=3.5]S[table-format=3.5]S[table-format=3.5]}
     & \multicolumn{8}{c}{\textbf{\ce{C6H9}-\ce{1-CN}}}\\
    \toprule
        && \multicolumn{3}{c}{Experimental} && \multicolumn{3}{c}{Theoretical} \\
        {Atom} && {$a$-coordinate} & {$b$-coordinate} & {$c$-coordinate} && {$a$-coordinate} & {$b$-coordinate} & {$c$-coordinate}\\
  \cline{1-1}
    \cline{3-5}
    \cline{7-9}
        && &  \\
        $C_\mathrm{1}$ && -0.601(3)\tnote{a,b} & 0.00(3)\tnote{c} & 0.00(7)\tnote{c} && -0.619 & -0.049 & 0.002\\
        $C_\mathrm{2}$ && 0.00(2)\tnote{c} & -1.210(1) & 0.10(2) && 0.048 & -1.210 & 0.098 \\
        $C_\mathrm{3}$ && 1.548(1) & -1.298(1) & 0.08(2) && 1.545 & -1.301 & 0.094 \\
        $C_\mathrm{4}$ && 2.2025(7) & 0.00(2)\tnote{c} & -0.370(4) && 2.214 & 0.004 & -0.354 \\
        $C_\mathrm{5}$ && 1.542(1) & 1.203(1) & 0.307(5) && 1.543 & 1.210 & 0.309 \\
        $C_\mathrm{6}$ && 0.00(3)\tnote{c} & 1.302(1) & -0.08(2) && 0.066 & 1.301 & -0.087 \\
        $C_\mathrm{7}$ && -2.0488(7) & -0.05(3) & -0.03(6) && -2.049 & -0.048 & -0.019 \\
        $N$ && -3.2073(5) & -0.04(4) & -0.04(4) && -3.204 & -0.005 & -0.041 \\
        \bottomrule
    \end{tabular}
    \begin{tablenotes}
        \item[a] Kraitchman coordinates are the result of a square root calculation and therefore the signs for each coordinate are taken from the calculated geometry.
        \item[b] Value in parentheses is the Costain error\cite{costain_further_1966}.
        \item[c] An imaginary value was calculated for this value and is reported as 0 within uncertainty.
    \end{tablenotes}
    \end{threeparttable}
\end{table*}

\begin{table*}[h!]
    \centering
    \caption{Experimentally determined Kraitchman coordinates of the heavy atoms of axial 4-cyanocyclohexene, {ax. \ce{C6H9}-\ce{4-CN}}, and equatorial 4-cyanocyclohexene, {eq. \ce{C6H9}-\ce{4-CN}}, compared to those calculated at the B3LYP-D3(BJ)/6-311G++(d,p) level of theory. Coordinates are given in Angstroms (\r{A}).}
    \label{tab:ax_eq_kraitchman_coords}
    \begin{threeparttable}
    \begin{tabular}{lcS[table-format=3.6]S[table-format=3.5]S[table-format=3.5]cS[table-format=3.5]S[table-format=3.5]S[table-format=3.5]}
     & \multicolumn{8}{c}{\textbf{axial {\ce{C6H9}-\ce{4-CN}}}}\\
    \toprule
        && \multicolumn{3}{c}{Experimental} && \multicolumn{3}{c}{Theoretical} \\
        {Atom} && {$a$-coordinate} & {$b$-coordinate} & {$c$-coordinate} && {$a$-coordinate} & {$b$-coordinate} & {$c$-coordinate}\\
  \cline{1-1}
    \cline{3-5}
    \cline{7-9}
        && &  \\
        $C_\mathrm{1}$ && -1.6054(9)\tnote{a,b} & -0.172(8) & 0.866(2) && -1.638 & -0.157 & 0.847\\
        $C_\mathrm{2}$ && -1.331(1) & 1.072(1) & 0.504(3) && -1.342 & 1.088 & 0.477 \\
        $C_\mathrm{3}$ && -0.311(5) & 1.435(1) & -0.554(3) && -0.311 & 1.428 & -0.568 \\
        $C_\mathrm{4}$ && 0.574(3) & 0.235(6) & -0.916(2) && 0.598 & 0.224 & -0.903 \\
        $C_\mathrm{5}$ && -0.260(6) & -1.031(1) & -1.076(1) && -0.249 & -1.059 & -1.049 \\
        $C_\mathrm{6}$ && -0.960(2) & -1.390(1) & 0.220(7) && -0.997 & -1.380 & 0.248 \\
        $C_\mathrm{7}$ && 1.6059(9) & 0.05(3) & 0.13(1) && 1.627 & 0.044 & 0.124 \\
        $N$ && 2.4092(6) & -0.11(1) & 0.938(2) && 2.434 & -0.111 & 0.934 \\
        \bottomrule
        & \\
    & \multicolumn{8}{c}{\textbf{equatorial {\ce{C6H9}-\ce{4-CN}}}}\\
    \toprule
        && \multicolumn{3}{c}{Experimental} && \multicolumn{3}{c}{Theoretical} \\
        {Atom} && {$a$-coordinate} & {$b$-coordinate} & {$c$-coordinate} && {$a$-coordinate} & {$b$-coordinate} & {$c$-coordinate}\\
  \cline{1-1}
    \cline{3-5}
    \cline{7-9}
        && &  \\
        $C_\mathrm{1}$ &&  -2.2474(7)\tnote{a,b} & -0.05(3) & -0.04(4) && -2.248 & -0.091 & -0.033 \\
        $C_\mathrm{2}$ && -1.562(1) & -1.233(1) & 0.08(2) && -1.567 & -1.232 & 0.073 \\
        $C_\mathrm{3}$ && 0.00(3)\tnote{c} & -1.304(1) & 0.10(2) && -0.064 & -1.309 & 0.109 \\
        $C_\mathrm{4}$ && 0.552(3) & 0.00(2)\tnote{c} & -0.365(4) && 0.574 & 0.014 & -0.365 \\
        $C_\mathrm{5}$ && -0.07(2) & 1.209(1) & 0.313(5) && -0.119 & 1.219 & 0.305 \\
        $C_\mathrm{6}$ && -1.590(1) & 1.265(1) & -0.10(2) && -1.596 & 1.265 & -0.095 \\
        $C_\mathrm{7}$ && 2.0093(8) & 0.02(9) & -0.12(1) && 2.014 & 0.019 & -0.116 \\
        $N$ && 3.1466(5) & 0.04(4) & 0.09(2) && 3.147 & 0.024 & 0.098 \\
        \bottomrule
    \end{tabular}
    \begin{tablenotes}
        \item[a] Kraitchman coordinates are the result of a square root calculation and therefore the signs for each coordinate are taken from the calculated geometry.
        \item[b] Value in parentheses is the Costain error\cite{costain_further_1966}.
        \item[c] An imaginary value was calculated for this value and is reported as 0 within uncertainty.
    \end{tablenotes}
    \end{threeparttable}
\end{table*}

\clearpage

\end{suppinfo}

\end{document}